\documentclass[%
 preprint,
 amsmath,amssymb,
 aps,
]{revtex4-2}

\usepackage{graphicx}
\usepackage{dcolumn}
\usepackage{bm}
\usepackage{braket,mathrsfs,amsmath}
\providecommand\JournalTitle[1]{#1}

\begin{document}


\title{Mitigating indistinguishability issues in photon pair sources by delayed-pump Intermodal Four Wave Mixing}

\author{Massimo Borghi}
\altaffiliation[Corresponding author:]{massimo.borghi@unipv.it}
\affiliation{%
 Nanoscience laboratory, Department of Physics, University of Trento, Via Sommarive 14, 38123, Trento, Italy
}%
\affiliation{%
 present address: Department of Physics, University of Pavia, Via Agostino Bassi, 6, 27100, Pavia, Italy
}%
\author{Lorenzo Pavesi}%
\affiliation{%
 Nanoscience laboratory, Department of Physics, University of Trento, Via Sommarive 14, 38123, Trento, Italy
}%




\date{\today}

\begin{abstract}
\noindent Large arrays of independent, pure and identical heralded single photon sources are ubiquitous in today's Noise Intermediate Scale Quantum devices (NISQ). In the race towards the development of increasingly ideal sources, delayed-pump Intermodal Four Wave Mixing (IFWM) in multimode waveguides has recently demonstrated record performances in all these metrics, becoming a benchmark for spontaneous sources in integrated optics. Despite this, fabrication imperfections still spoil the spectral indistinguishability of photon pairs from independent sources. Here we show that by tapering the width of the waveguide and by controlling the delay between the pump pulses, we add spectral tunability to the source while still inheriting all the record metrics of the IFWM scheme. This feature is used to recover spectral indistinuishability in presence of fabrication errors. Under realistic tolerances on the waveguide dimensions, we predict $>99.5\%$ indistinguishability between independent sources on the same chip, and a maximum degradation of the Heralded Hong Ou Mandel visibility $<0.35\%$.
\end{abstract}

\maketitle


\section{Introduction}
\noindent Spontaneous sources of photon pairs are primary resources in emerging large scale NISQ architectures, especially those based on integrated optics \cite{caspani2017integrated,wang2020integrated}. Through repeated application of heralding, large arrays of sources can be used to deterministically prepare many independent photons, which constitutes an important substrate for quantum information processing \cite{bonneau2015effect,collins2013integrated}. Their quality influences the ultimate computational power of the hardware, and limits the effective size of resources which are available for quantum algorithms \cite{renema2018efficient,shchesnovich2014sufficient,sparrow2017quantum}.  Two of the most relevant metrics are the purity and the indistinguishability of the heralded states \cite{signorini2020chip}. In essence, they bound the visibility of multiphoton interference, which lies at the heart of protocols, algorithms and building blocks for quantum computation and quantum information. Examples include  scattershot \cite{paesani2019generation} and gaussian boson sampling \cite{arrazola2021quantum}, preparation of cluster states \cite{vigliar2021error}, realization of entangling gates \cite{adcock2018hard,adcock2019programmable} and state teleportation \cite{llewellyn2020chip}. Several devices and methods have been developed to herald photons in pure states, characterized by a single and well defined spectral-temporal mode. These span from phase matching engineering  \cite{graffitti2018independent}, pump manipulation \cite{burridge2020high}, selective control of the quality factor in microresonators  \cite{liu2020high} and spectral filtering \cite{blay2017effects}.  Even if the purity can be improved from a clever design of the device, the indistinguishability relies exclusively on the fabrication uniformity of the array of sources. To date, even state of the art lithographic techniques can not guarantee sufficient uniformity levels, and errors must be compensated in post-fabrication. Indeed, the thickness uniformity of the silicon waveguide layer (long range disorder) has a rms value of $3-4$ nm, while at die level ($\sim$ cm$^{2}$ size, short range disorder), the uniformity in the waveguide width  has an rms value $<10\,\textrm{nm}$ \cite{siew2021review}. \\
\noindent Independent sources based on microresonators can be made indistinguishable by aligning and locking their resonance wavelengths through thermo optic tuning \cite{silverstone2015qubit,arrazola2021quantum}. However, this method does not compensate slightly differences in the Free Spectral Range (FSR) or in the cavity linewidth, which are especially relevant for resonators of high quality factor. Waveguide sources without phase matching engineering emit photons in a broad spectral interval, and off or on-chip filters are used to increase their purity at the expense of reducing the heralding efficiency \cite{meyer2017limits}. Therefore, the indistinguishability depends on the fabrication uniformity of the filters. In general, waveguide sources of spectrally uncorrelated photon pairs are not easily reconfigurable. Small tuning ranges can be obtained by heating the whole chip \cite{kumar2013spectrally}, while wider variations require to modify the pump wavelength \cite{jin2013widely}. Other techniques aim to erase the spectral distinguishability only after that the pair is generated. This can be achieved in materials with a strong second order nonlinearity by electro-optic frequency shearing \cite{zhu2021spectral}, or in third order materials by Four Wave Mixing Bragg Scattering \cite{li2016efficient}. \\
\noindent In this work, we propose and validate the design of a waveguide source which emits highly pure and spectrally tunable photons without spectral filtering. This is achieved through delayed-pump Intermodal Four Wave Mixing, a scheme recently reported on the SOI platform and which showed a record heralded Hong Ou Mandel (HHOM) visibility of $96\%$ between independent sources \cite{paesani2020near}. In contrast to the original work, we introduce an adiabatic change of the waveguide width along the propagation direction, and we tune the relative delay between the two pumps to reconfigure the phase matching wavelength of the emitted photons. The delay determines the point where the pump pulses overlap, which in turn selects the segment of the waveguide where pair generation occurs. Since the Signal/Idler frequencies depend on the waveguide cross-section, the delay reconfigures the generation wavelengths of the photon pair. We numerically investigate how this feature can be used to mitigate the distinguishability issues between different sources which arise from fabrication imperfections. We consider errors on both the waveguide width and height, focusing on realistic ranges provided by commercial foundries. We show that an indistinguishability level $>95\%$ can be guaranteed up to height differences of $4.3\,\textrm{nm}$, and for width differences greater than $100$ nm. The HHOM visibility is shown to degrade by less than $0.35\%$ from its value in two identical sources for devices on the same chip. In all the considered cases, the spectral tunability allows to dramatically improve the visibility of both Reverse (RHOM) and Heralded Hong Ou Mandel interference. 
We also prove that the principal source metrics and the spectral tunability are not degraded by the Self and the Cross Phase Modulation induced by the pump on the Signal and the Idler photon.

\section{Principle of operation and theory}
\label{sec:principle}
\noindent In spontaneous IFWM, photons from two bright Pump fields (labeled as $p_1$ and $p_2$) annihilate to produce Signal ($s$) and Idler ($i$) photon pairs propagating in the different transverse mode orders of a multimode waveguide. The generation process occurs within narrow frequency ranges located at large spectral distances from the pump wavelength, where phase matching is satisfied \cite{signorini2018intermodal}. By denoting the wavevectors of the  fields as  $(k_{p_1},k_{p_2},k_s,k_i)$, and their central wavelengths as $(\bar{\lambda}_{p1},\bar{\lambda}_{p2},\bar{\lambda}_s,\bar{\lambda}_i)$, this condition implies that  $k_{p1}(\bar{\lambda}_{p1})+k_{p2}(\bar{\lambda}_{p2})=k_s(\bar{\lambda}_s)+k_i(\bar{\lambda}_i)$. The great flexibility offered by the choice of the modal combination and by the waveguide cross section has been exploited to tune the emission wavelengths from the Near Infrared to the Mid-Infrared  \cite{signorini2018intermodal,signorini2021silicon} range. At the same time, the narrow generation bandwidth and the different group velocities of the modes can be exploited to engineer the emission of spectrally uncorrelated photon pairs. Within this framework, we revisit the configuration described in \cite{paesani2020near}, where IFWM is demonstrated on a $220$ nm thick SOI waveguide.
.
\subsection{Tuning the phase matching wavelengths with the waveguide width}
\begin{figure}[t!]
\centering\includegraphics[scale = 0.78]{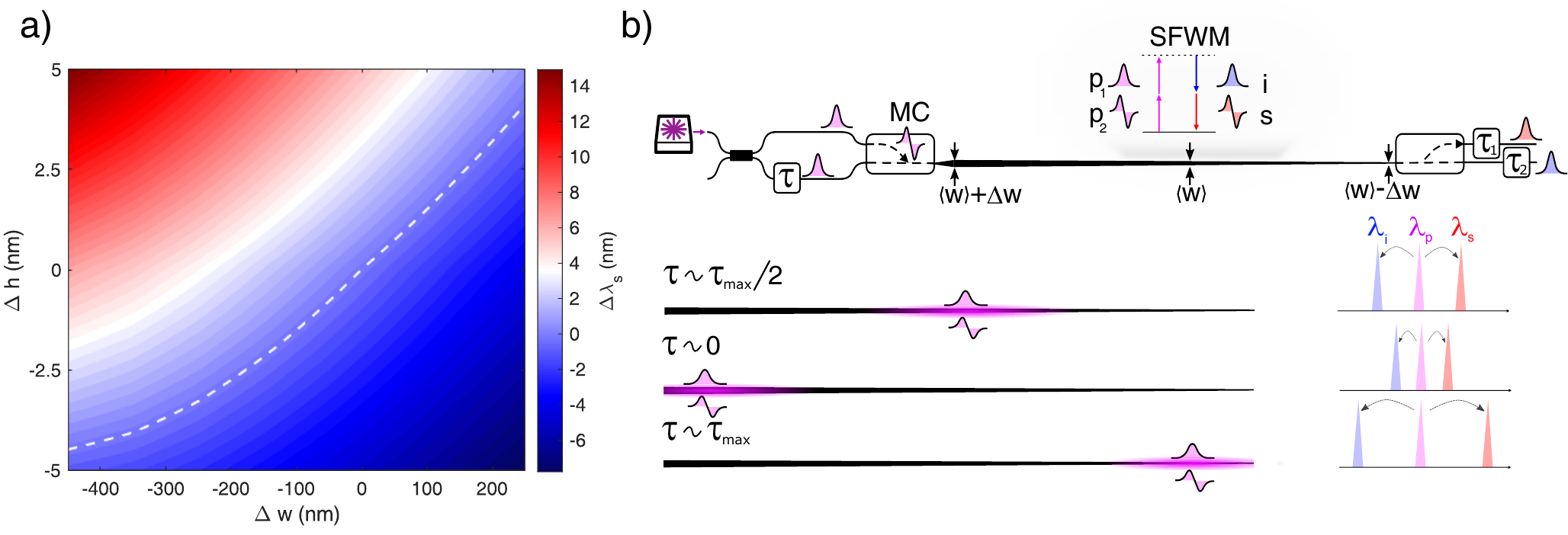}
\caption{(a) Variation of the phase matching wavelength of the Signal $\Delta\lambda_s$ with respect to changes in the width ($\Delta w$) and height ($\Delta h$) of the reference waveguide cross-section $2.25\,\times 0.22\,\mu\textrm{m}$. The white dashed line follows $\Delta\lambda_s=0$. (b) Top: sketch of the source, with indicated the relevant components and parameters (MC = Mode Converter). Bottom: principle of operation of the tunable source. In order to tune the phase matching wavelengths of the Signal ($\lambda_s$) and of the Idler ($\lambda_i$)  (right sketch), the two pump pulses are delayed by a variable amount of time $\tau$, making them to overlap in different positions (magenta color) along the waveguide. At each delay, a different waveguide width is sensed, and the phase matching wavelengths change according to panel (a). The delays $\tau_1$ and $\tau_2$ are respectively applied to the Signal and the Idler photon to control their arrival time. \label{Fig_1}}
\end{figure}

\noindent We use a Pump pulse of gaussian shape with a Full Width at Half Maximum (FWHM) duration of $T_0=0.8\,\textrm{ps}$, a repetition rate of $50\,\textrm{MHz}$ and a wavelength of $1550\,\textrm{nm}$. This is coupled in a coherent superposition of the two lowest order Transverse Magnetic (TM) modes (TM0 and TM1 mode), with a relative delay $\tau$ between them. From now on, we will refer to the faster and delayed pulse in the TM0 mode as the pump $1$, while the pulse in the TM1 mode as the pump $2$. The multimode waveguide has a width of $\langle w \rangle = 2.25\,\mu\textrm{m}$ and a length of $L=1.5\,\textrm{cm}$. Signal and Idler photons are generated in the TM1 and TM0 modes respectively at the wavelengths $\bar{\lambda}_{s}=1581.4\,\textrm{nm}$ and $\Bar{\lambda}_{i}=1519.9\,\textrm{nm}$. Geometrical variations with respect to this reference configuration lead to a shift $\Delta\lambda_{s(i)}$ of their phase matching wavelengths. This is shown in Fig.\ref{Fig_1}(a), in which $\Delta\lambda_s$ is plotted as a function of the deviation $\Delta w$ and $\Delta h$ in the waveguide width and height. Due to the remarked sensitivity of TM modes with the latter, we have that $\frac{d\lambda_s}{d\Delta h} \sim 1$ while $\frac{d\lambda_s}{d\Delta w} \sim -0.015$. Despite this, changing the waveguide width is easier than locally varying the thickness of the silicon device layer, so we can adjust $\Delta w$ to tailor the emission wavelength of the source. We exploit two key characteristics of IFWM to realize a single device which can be reconfigured. The first is that due to the temporal walk-off between the pump pulses, the position $z = L_{\textup{match}}$ along the waveguide where the pair generation probability is maximum depends on the delay $\tau$. This is given by $L_{\textup{match}}=\tau \left ( \frac{1}{v_{p1}}-\frac{1}{v_{p2}} \right )$ (here, $v_{p1(2)}$ is the group velocity of pump $1(2)$), which is the coordinate where the two pump pulses overlap (see Appendix C).
The effective width which determines the phase matching wavelengths corresponds to the local waveguide width $w(z)$ at position $z = L_{\textup{match}}$. 
 The second feature which we exploit is the fact that by letting $w$ to vary along the propagation direction, the effective width where pair generation occurs can be controlled with $\tau$. As a consequence, the generation wavelengths can be continuously tuned, as shown in Fig.\ref{Fig_1}(a). We focus on  the configuration shown in Fig.\ref{Fig_1}(b), where the width of the waveguide is linearly tapered from $w(z=0) = \langle w \rangle+ \Delta w$ to $w(z = L) = \langle w \rangle- \Delta w$, with $\Delta w \ge0$. We define $\tau_{\textup{max}}=4.9\,\textrm{ps}$ as the delay which makes the two pump pulses to overlap at the end of the waveguide. In Fig.\ref{Fig_1}(b) we analyze three extremal cases. When $\tau=0$, the maximum pump overlap occurs at $\langle w \rangle+\Delta w$, and according to Fig.\ref{Fig_1}(a), $\Delta\lambda_{s}<0$, i.e., pairs are generated at wavelengths closer to the one of the pump. When $\tau = \frac{\tau_{\textup{max}}}{2}$, the overlap is maximum at $\langle w \rangle$, and the phase matching wavelengths are not changed with respect to the case $\Delta w =0$. When $\tau = \tau_{\textup{max}}$, the pump pulses catch at the narrower end of the waveguide, and photon pairs are generated at larger spectral detunings with respect to the pump wavelength. As long as $L$ exceeds the walk-off length between the Pump pulses, and that the choice of $\tau$ allows a complete progression of one pulse over the other, the generation bandwidth and the efficiency remains constant. In the next section, we quantitatively evaluate $\Delta\lambda_{s(i)}$ as a function of $\tau$ and $\Delta w$, focusing on how the pair generation probability and the purity of the heralded single photon states are affected.

\subsection{Theory of photon pair generation in the tapered source}
\noindent The spectral (temporal) properties of photon pairs are characterized their Joint Spectral (Temporal) Amplitude (JSA/JTA), and most of the source metrics can be derived from this function \cite{christ2011probing}. We then focus on the derivation of the JSA/JTA, taking into account the multiple spatial modes in the FWM process, the delayed pump configuration, the varying waveguide width along the propagation direction and the effects of SPM and XPM between the pumps and the Signal/Idler photons. The electric fields of the two pumps are treated classically and are expressed as  \cite{koefoed2017effects}:
\begin{equation}
    \mathbf{E}_{p1(2)}(z,t) = \frac{1}{2}\mathbf{e}F_{p1(2)}(x,y) \sqrt{\frac{2}{n\epsilon_0 c}}\Tilde{A}_{p1(2)}(z,t)e^{i(\Bar{\beta}_{p1(2)}z-\Bar{\omega}_{p1(2)}t)} + \textrm{c.c},
    \label{eq:1}
\end{equation}
where $\mathbf{e}$ is the unit vector of polarization, $F_p$ is the transverse mode profile (normalized such that $\int \left |F_p(x,y)\right |^2dxdy = 1$), $n$ is the refractive index of the waveguide core, $\bar{\beta}_{p1(2)}$ and $\bar{\omega}_{p1(2)}$ the central wavevector and frequency of the fields, and $\Tilde{A}_{p1(2)}$ a slowly varying envelope function. The power carried by the field in Eq.(\ref{eq:1}) is $P_{p1(2)}=|\Tilde{A}_{p1(2)}|^2$, as can be verified by integrating the Poyinting vector $\mathbf{S}=\frac{\mathbf{E}\times\mathbf{B}}{\mu_0}$ across the waveguide cross section. The two pump envelopes are temporally delayed gaussians, and are defined in Appendix C. 
The Signal and the Idler fields are quantized as:
\begin{equation}
    \mathbf{E}_{s(i)}(z,t) = \frac{1}{2}\mathbf{e}F_{s(i)}(x,y)\frac{1}{2\pi}\int \sqrt{\frac{2\hbar\omega}{n\epsilon_0 c}}\Tilde{a}_{s(i)}(z,\omega)e^{-i\omega t}e^{i(\Bar{\beta}_{s(i)}z-\Bar{\omega}_{s(i)}t)}d\omega + \textrm{c.c},
    \label{eq:2}
\end{equation}
where $\Tilde{a}_{s(i)}(z,\omega)$ represents the Fourier Transform of the slowly varying annihilation operator $\Tilde{a}_{s(i)}(z,t)$ for the Signal(Idler) photon. It is possible to formally derive the propagation equation for $\mathbf{E}_{p1(2)}$ in a fully quantum mechanical framework by treating $\Tilde{A}_{p1(2)}$ as an operator and by using the Heisenberg equation of motion. However, we anticipate the result of the classical regime, in line with the fact that the field in Eq.(\ref{eq:1}) is not quantized. This is given by the well known set of coupled Nonlinear Schrodinger equations (NLSE) \cite{agrawal1989temporal}:
\begin{equation}
    \frac{\partial \Tilde{A}_{p1}}{\partial z} = \left ( -\frac{\alpha_{p1}}{2}+i\Delta\beta_{p1}(z)-\frac{i}{2L_{D_{p1}}}\frac{\partial^2}{\partial T^2} \right )\Tilde{A}_{p1}+i\left (2\gamma_{1122}|\Tilde{A}_{p2}|^2+ \gamma_{1111}|\Tilde{A}_{p1}|^2\right )\Tilde{A}_{p1},
    \label{eq:3}
\end{equation}
\begin{equation}
    \frac{\partial \Tilde{A}_{p2}}{\partial z} = \left ( -\frac{\alpha_{p2}}{2}+i\Delta\beta_{p2}(z)-\frac{i}{2L_{D_{p2}}}\frac{\partial^2}{\partial T^2} -\frac{1}{L_{w_p}}\frac{\partial}{\partial T}\right )\Tilde{A}_{p2}+i\left (2\gamma_{2211}|\Tilde{A}_{p1}|^2+ \gamma_{2222}|\Tilde{A}_{p2}|^2\right )\Tilde{A}_{p2},
     \label{eq:4}
\end{equation}
where the dimensionless time $T= \left (t-\frac{z}{v_{g1}} \right )/T_0$ refers to a reference frame moving at the group velocity of pump 1. The definition of the parameters can be found in Appendix A. The second term on the right hand side of Eqs.(\ref{eq:3},\ref{eq:4}) is defined as $\Delta\beta_{p}=\beta_{p}(z,\bar{\omega}_{p})-\bar{\beta}_{p}$, and accounts for the varying waveguide width along the propagation direction. We numerically integrated this set of equations using a third order, symmetrized Split-Step Fourier method (SSFM) \cite{agrawal2000nonlinear}. To obtain a similar set of equations for the Signal and the Idler field operators, we use the Heisenberg equation of motion generated by the momentum operator $M(z,t)$, which is $-i\hbar \frac{dO}{dz}=[O,M]$ \cite{huttner1990quantum}, where $O$ is any operator in the Heisenberg picture. The total momentum can be written as $M = M_L+M_{\textrm{SPM}}+M_{\textrm{XPM}}+M_{\textrm{FWM}}$, which is the sum of the linear, the SPM, the XPM and the FWM induced momentum \cite{sinclair2016effect}, and whose expressions can be found in Appendix A. We then move in the interaction picture and split the total momentum into $M=M_0+M_{\textrm{FWM}}$, where all the trivial evolution is generated by $M_0=M_L+M_{\textrm{SPM}}+M_{\textrm{XPM}}$.  The pair generation process is described by the interaction momentum $M_{\textrm{FWM}}$. Using the expressions for $M_L$, $M_{\textrm{SPM}}$ and $M_{\textrm{XPM}}$ provided in Appendix A, and the equal position commutation relation $[\Tilde{a}_{s(i)}(z,t),\Tilde{a}_{s(i)}^{\dagger}(z,t')]=\delta(t-t')$ \cite{koefoed2017effects}, we get \cite{bell2015effects}:
\begin{equation}
     \label{eq:5}
\begin{aligned}
    \frac{\partial \Tilde{a}_{s(i)}}{\partial z}  = & \left ( -\frac{\alpha_{s(i)}}{2}+i\Delta\beta_{s(i)}(z)-\frac{i}{2L_{D_{s(i)}}}\frac{\partial^2}{\partial T^2} -\frac{1}{L_{w_{s(i)}}}\frac{\partial}{\partial T}\right )\Tilde{a}_{s(i)}+2i\left (\gamma_{11s(i)s(i)}|\Tilde{A}_{p1}|^2 \right. + \\
    {} & + \left. \gamma_{22s(i)s(i)}|\Tilde{A}_{p2}|^2\right )\Tilde{a}_{s(i)},
     \end{aligned}
\end{equation}    
where we have neglected the XPM and the SPM of the Signal and the Idler fields. It is worth to note that losses have been phenomenologically introduced by the linear loss coefficients $\alpha_{s(i)}$. Losses spoil the photon number correlation between the Signal and the Idler photon in the two-mode squeezed state generated by $M_{\textrm{FWM}}$, which could be accounted by introducing a reservoir of loss modes that is coupled to the Signal/Idler fields \cite{helt2015spontaneous}. Beside that, the simultaneous presence of squeezing and loss differs from the case where the two effects separately act  \cite{helt2015spontaneous}. However, the latter well approximates the case of IFWM, since the interaction length is small compared to the one of the waveguide, and losses can be assumed to be all lumped after pair generation. Provided that we restrict our attention to the low squeezing regime of single pair generation, the loss term in Eq.(\ref{eq:5}) simply scales the pair generation probability by a factor $\alpha_s\alpha_i$, and does not contribute to modify the shape of the JSA. The state $\ket{\Psi(z)}$ of the Signal and the Idler photon, lying in vacuum at $z=0$, evolves as $-i\hbar \frac{d\ket{\Psi}}{dz}=M_{\textrm{FWM}}\ket{\Psi}$ \cite{sinclair2016effect}, and its solution can be formally written in terms of a space propagator $\ket{\Psi(z)}=U(z,0)\ket{\Psi(0)}$ \cite{koefoed2017effects}. In the regime of single pair generation, this is given by $U(z,0)= I+\frac{i}{\hbar}\int_{0}^{z}M_{\textrm{FWM}}(z')dz'$, where $I$ denotes the identity operator. From the two-photon state, we can define the joint amplitude probability $\Phi(T_s,T_i,z)$ of detecting, at position $z$, the Signal photon at time $T_s$ and the Idler photon at time $T_i$,  as $\Phi(T_s,T_i,z)=\langle \Tilde{a}_s(T_s,z)\Tilde{a}_i(T_s,z)U(z,0) \rangle$, where the expectation value is evaluated on vacuum. When $\Phi$ is normalized such that $\int |\Phi|^2dT_sdT_i=1$, this coincides with the definiton of the JTA \cite{bell2015effects}. In the rest of the paper, we will refer to $\Phi(T_s,T_i,z)$ as the JTA without distinction. The JSA $\Phi(\omega'_s,\omega'_i,z)$, expressed in the dimensionless frequencies $\omega'_{s(i)}=(\omega_{s(i)}-\bar{\omega}_{s(i)})T_0$, is related to $\Phi(T_s,T_i,z)$ by a two-dimensional Fourier Transform \cite{sinclair2016effect}. Following the derivation detailed in Appendix B, and similarly reported in \cite{koefoed2019complete}, we can write a propagation equation for the JTA. By expressing the latter as $\Phi = \Tilde{\Phi}(T_s,T_i,z)e^{i\Theta_{si}(z)}$, where $\Theta_{si} = \int_{0}^{z}(\Delta\beta_{s}(z')+\Delta\beta_{i}(z'))dz'$, the function $\Tilde{\Phi}$ obeys the equation:
\begin{equation}
    \frac{\partial \Tilde{\Phi}(T_s,T_i,z)}{\partial z} = \left ( L_s+L_i+N_s+N_i\right )\Tilde{\Phi}(T_s,T_i,z)+S(T_s,T_i,z), \label{eq:6}
\end{equation}
where the operators $L_{s(i)}$, $N_{s(i)}$ and the driving term $S$ are defined as:
\begin{equation}
\label{eq:7}
    \begin{aligned}
        L_{s(i)} = & -\frac{\alpha_{s(i)}}{2}-\frac{1}{L_{w_{s(i)}}}\frac{\partial}{\partial T_{s(i)}}-\frac{i}{2L_{D_{s(i)}}}\frac{\partial^2}{\partial T_{s(i)}^2}, \\
        N_{s(i)} = & \,2i(\gamma_{11s(i)s(i)}|\Tilde{A}_{p1}(z,T_{s(i)})|^2+\gamma_{22s(i)s(i)}|\Tilde{A}_{p2}(z,T_{s(i)})|^2), \\
        S = & \gamma_{p1p2si}e^{i\Theta(z)}\int G(\omega_s',\omega_i',z)e^{-i(T_s\omega_s'+T_i\omega_i')}d\omega_s' d\omega_i', \\
        G(\omega_s',\omega_i',z) = & i\int \bar{\mathscr{A}}_{p1}(x,z) \bar{\mathscr{A}}_{p2}(\omega_s'+\omega_i'-x,z) dx, \\
        \Theta (z) = & \int_{0}^{z}(\Delta\beta_{p1}(z')+\Delta\beta_{p2}(z')-\Delta\beta_s(z')-\Delta\beta_i(z'))dz',
    \end{aligned}
\end{equation}
where we wrote the Fourier Transform of $\Tilde{A}_{p1(2)}(z,T)$ as $\bar{\mathscr{A}}_{p1(2)}(z)\exp{i\left ( \int_0^z \Delta\beta_{p1(2)}(z')dz'\right )}$ to factor out the accumulated phase due to the tapering. Equation \ref{eq:6} has the same structure of a two-dimensional NLSE in the dimensionless time variables $(T_s,T_i)$, with the inclusion of an external driving term $S$. 
\section{Analysis of the source performance}
\label{sec:source_metrics}
\begin{figure}[t!]
\centering\includegraphics[scale = 0.72]{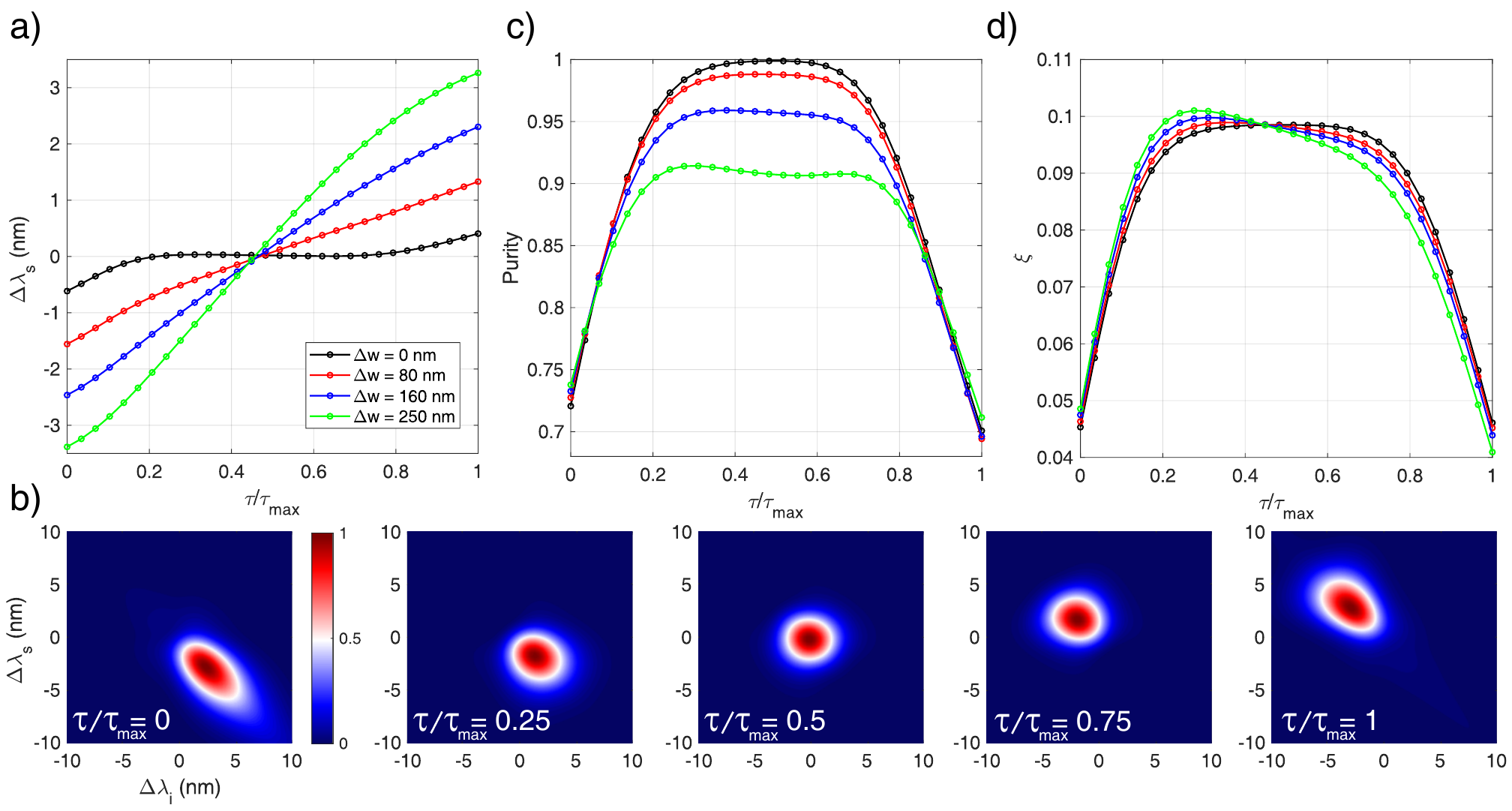}
\caption{(a) Relative shift of the average Signal wavelength as a function of the time delay $\tau$ between the pumps. Different curves refer to different tapering amplitudes $\Delta w$. (b) JSA of the photon pair source for different choices of the delay $\tau$. The tapering amplitude is fixed to $\Delta w = 0.25\,\mu\textrm{m}$. (c) Purity of the heralded photon states as a function of the delay $\tau$ and for different values of $\Delta w$. (d) Same as in (c), but relative to the pair generation probability $\xi$.  \label{Fig_2}}
\end{figure}

\noindent Using the third order SSFM developed in \cite{koefoed2017effects,koefoed2019complete}, we numerically integrated Eq.\ref{eq:6} to calculate the JTA and the JSA for different tapering amplitudes $\Delta w$ and for different delays $\tau$. The average pump power is set to $1$ mW, and is equally distributed between the TM0 and the TM1 modes. The mean wavelength shift $\Delta\lambda_s$ of the Signal photon, calculated from the JSA, is shown in Fig.\ref{Fig_2}(a), while the related JSAs (plotted here only for $\Delta w = 0.25\,\mu\textrm{m}$) are shown in Fig.\ref{Fig_2}(b). For a fixed value of $\Delta w$, the phase matching wavelengths are continuously tuned with $\tau$. The trends follow the one indicated in Fig.\ref{Fig_1}(a,b), where the spectral separation of the Signal/Idler wavelengths monotonically increases as $\tau\rightarrow \tau_{\textrm{max}}$. The maximum tuning range depends on $\Delta w$, and increases from $\Delta\lambda_s^{\textrm{max}} =\Delta\lambda_s(\tau = \tau_{\textrm{max}}) -\Delta\lambda_s(\tau = 0)\sim 2$ nm for $\Delta w = 0.08\,\mu\textrm{m}$  to $\Delta\lambda_s^{\textrm{max}}\sim 6.5$ nm for $\Delta w = 0.25\,\mu\textrm{m}$. Except for the extremal cases $\tau = \{0,\tau_{\textrm{max}}\}$, the JSA maintains an almost perfect circular shape. The generation bandwidth does not increase with $\Delta w$, which is an exclusive property of the delayed-pump IFWM scheme. If the tapering angle $2\Delta w/L$ is kept shallow, the local waveguide width does not appreciably change along the interaction length, and the generation bandwidth remains constant. The high purity of the Signal and the Idler photon is shown in Fig.\ref{Fig_2}(c). With respect to the a straight waveguide ($\Delta w=0$), for which the purity is maximum and equal to $\mathscr{P}=0.998$ at $\tau=0.5\tau_{\textrm{max}}$, this only decreases  to $\mathscr{P}=0.98$ for $\Delta w=0.08\,\mu\textrm{m}$ and to $\mathscr{P}=0.91$ for $\Delta w = 0.25\,\mu\textrm{m}$. The pair generation probability $\xi$ is almost not affected by $\Delta w$. As $\tau \rightarrow \tau_{\textrm{max}}$, the sensed effective area becomes smaller, but this does not improve the FWM strength since the two pump pulses accumulate more losses before overlapping at the narrower end of the waveguide. 
\begin{figure}[t!]
\centering\includegraphics[scale = 0.60]{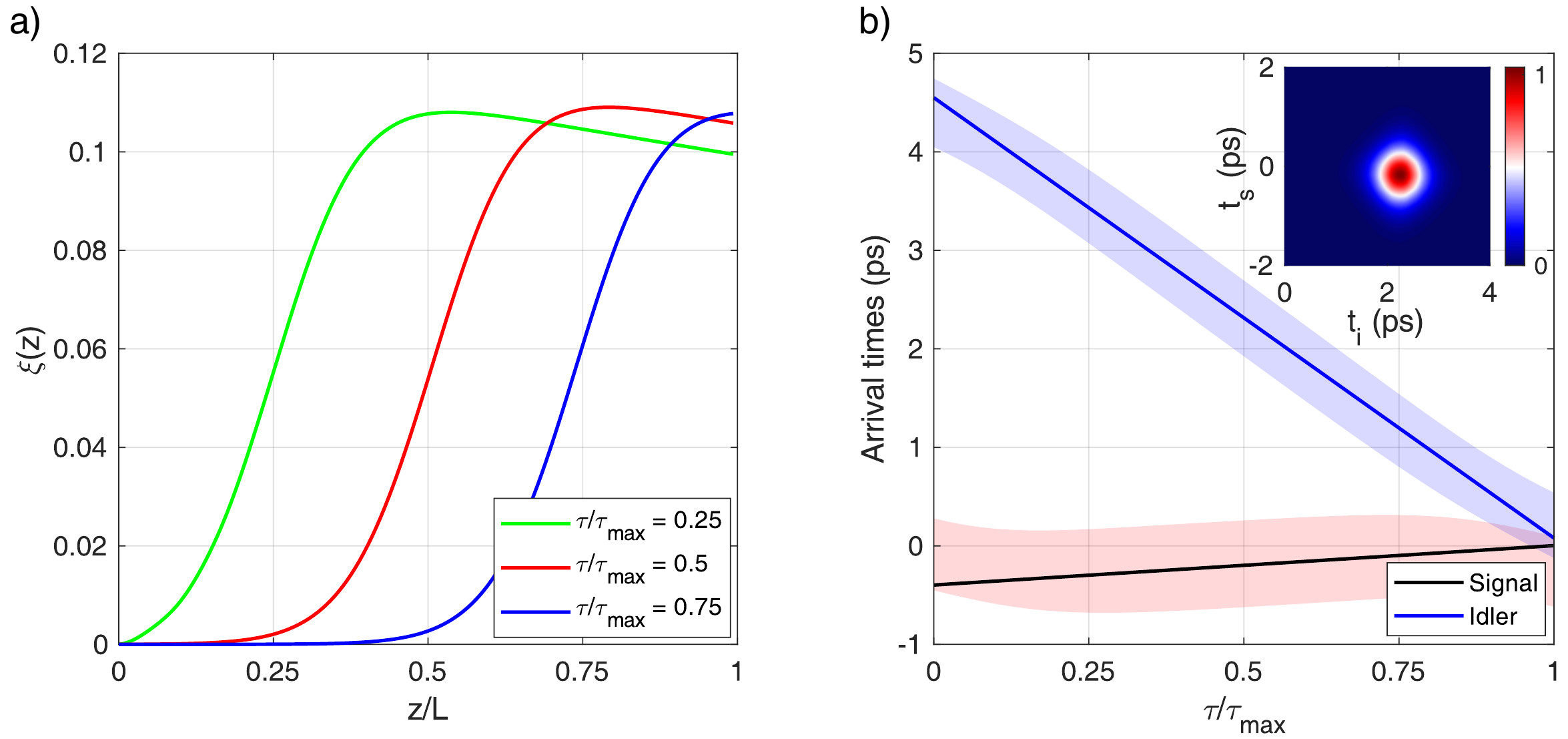}
\caption{(a) Cumulative photon pair generation probability $\xi$ as a function of the position $z$ along the waveguide. Different curves refer to different choices of the relative pump delay $\tau$. (b) Average arrival times of the Signal (black) and of the Idler (blue) photon at the end of the waveguide, calculated using Eq.(\ref{eq:9}). These values are relative to the arrival time of the pump pulse which lies in the TM0 mode. The inset shows an example of the JTA for $\frac{\tau}{\tau_{\textrm{max}}}=0.5$, from which the mean $\langle T_{s(i)}\rangle$ and the standard deviation $\sigma_{T_{s(i)}}$ on the arrival times are extracted. These are plotted as shaded regions ($\langle T_{s(i)} \rangle \pm \sigma_{T_{s(i)}}$). \label{Fig_3}}
\end{figure}
\noindent The source metrics discussed so far refer to the properties of the Signal/Idler pair at the end of the waveguide, but they do not offer a physical insight into the evolution of the two-photon state as it propagates along the source. One of the strengths of Eq.(\ref{eq:6}) is to provide a natural framework to track evolution of any metric along the waveguide. As an example, the accumulated pair generation probability $\xi(z)=\int |\Phi(T_s,T_i,z)|^2 dT_sdT_i$, from the beginning of the waveguide to position $z$, can be computed starting from Eq.(\ref{eq:6}) as:
\begin{equation}
    \xi(z) = \int\left ( \int_{0}^{z} \left (-(\alpha_s + \alpha_i)|\Tilde{\Phi}(T_s,T_i,z')|^2 +2\mathscr{R} \left [ S \Tilde{\Phi(T_s,T_i,z')}^*\right ] \right )dz' \right )dT_sdT_i,
    \label{eq:8}
\end{equation}
where $\mathscr{R}$ denotes the real part and we have used the fact that, from  Eq.\ref{eq:7}, $L_{s(i)}^{\dagger} = -L_{s(i)}+\alpha_{s(i)}$ and $N_{s(i)}^{\dagger} = -N_{s(i)}$. In Fig.\ref{Fig_3}(a), we plot $\xi$ as a function of $z$ for $\frac{\tau}{\tau_{\textrm{max}}} = \{0.25,0.5,0.75\}$ and $\Delta w = 0.1\,\mu\textrm{m}$. The essence of IFWM emerges from these curves. The generation probability is approximately zero until $\frac{z}{L}\sim \frac{\tau}{\tau_{\textrm{max}}}$, which is the point where the two pump pulses match. Then, the value of $\xi$ smoothly grows from the $5\%$ to the $95\%$ of its maximum in a length of $L_{\textrm{grow}}\sim0.36L$. Then, the cumulative generation probability saturates since the pump pulses lose their spatial overlap, and after that it exponentially decays due to the propagation losses. As shown in Appendix C, the function $\xi(z)$ can be approximated by an \emph{erf} function, which implies that its derivative, representing the pair generation probability per unit length, is a gaussian peaked at $\frac{z}{L}=\frac{\tau}{\tau_{\textrm{max}}}$. Its FWHM $\Delta_z$ can be assessed from $\Delta_z = \frac{\ln(2)}{2}L_{\textrm{grow}}\sim 0.21L$. This value is very close to the the approximated analytic result found in Appendix C, which is $\Delta_z=\sqrt{2}L_{w_p}=0.20L$. For $\Delta w=0.1\,\mu\textrm{m}$, the width of the waveguide changes by $\frac{2\Delta w\Delta_z}{L}\sim 0.06\,\mu\textrm{m}$ along $\Delta_z$, which is the $\sim 2.5\%$ of the waveguide width. 

\noindent Since the pair generation process is well localized in space, so they have to be the arrival times of the Signal and the Idler photon at the end of the waveguide. As shown from the JTA in the inset of Fig.\ref{Fig_3}(b), photons are generated in a well defined gaussian temporal wavepacket, whose  size is of the same order of the pump duration ($0.8$ ps). From the JTA, the mean and the standard deviation on the Signal/Idler arrival times are calculated, which are shown in Fig.\ref{Fig_3}(b)  (shaded regions) as a function of $\tau$ and for $\Delta w =0$. These values are relative to the arrival time of the faster pump pulse, in accordance to the fact that Eq.(\ref{eq:7}) is expressed in a moving reference frame. The arrival times can be analytically predicted by assuming that the pair is generated at the position where the two pump pulses have their maximum overlap, which for a delay $\tau$ occurs at $z=L_{\textrm{match}} = \frac{\tau T_0}{L_{w_p}}$. From $z=L_{\textrm{match}}$, the time required for the Signal (Idler) photon to reach the end of the waveguide is $\frac{L-L_{\textrm{match}}}{v_{s(i)}}$, from which is easy to show that the arrival times $\mathscr{T}_{s(i)}$ are given by:
\begin{equation}
    \mathscr{T}_{s(i)} =  \frac{1}{|L_{w_{s(i)}}|} \left (\pm L_{w_p} \tau \mp T_0 L \right ),
    \label{eq:9}
\end{equation}
where the $+$ sign is used for the Signal. The solid lines in Fig.\ref{Fig_3}(b), obained from Eq. (\ref{eq:9}), show a good agreement with the arrival times calculated from the JTA. 
\section{Mitigating indistinguishability issues in two photon interference}
\noindent We now exploit the tunability of the source to mitigate the indistinguishability issues which arise from fabrication imperfections in indepedent devices. 
Suppose to have two sources, labelled $1$ and $2$, which can either lie on the same die or on two different chips. 
 In general, due to fabrication imperfections, they will have a different cross-section and JSA, which will compromise their capability to interfere. We can try to recover their spectral indistinguishability by respectively applying pump delays $\tau_1$ and $\tau_2$ to the two sources in order to overlap their Signal/Idler spectra. Unfortunately, as shown in Fig.\ref{Fig_3}(b), whenever $\tau_1\neq\tau_2$, the Signal(Idler) photons will arrive at the end of the waveguide at the different times $\tau_{s1}(\tau_{i1})$ and $\tau_{s2}(\tau_{i2})$. In other terms, they will be spectrally indistinguishable but temporarily distinguishable. In order to erase the temporal information, additional delay stages have to be placed at the end of the waveguide, which make $\tau_{s1}(\tau_{i1})=\tau_{s2}(\tau_{i2})$. To this purpose, the same component used to delay the pump pulses can be implemented, as shown in Fig.\ref{Fig_1}(b). 
\begin{figure}[t!]
\centering\includegraphics[scale = 0.8]{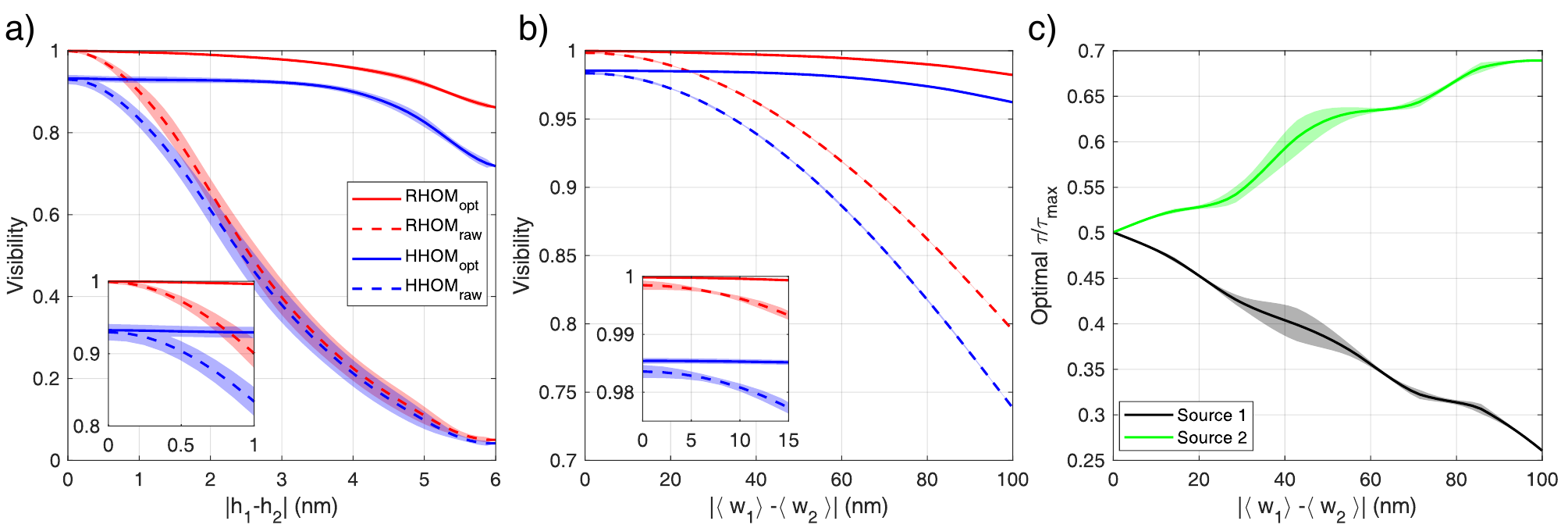}
\caption{(a) RHOM and HHOM  visibilities  as a function of the height difference of the two waveguide sources in the delay-optimized ($\textrm{RHOM}_{\textrm{opt}}$, $\textrm{HHOM}_{\textrm{opt}}$) and not optimized ($\textrm{RHOM}_{\textrm{raw}}$,$\textrm{HHOM}_{\textrm{raw}}$) cases. The inset shows a detail of the region $|h_1-h_2|\leq 1$. Visibilities $\mathscr{V}$ are calculated in the two dimensional grid spanned by $h_1$ and $h_2$, which are stored in the matrix $\mathscr{V}_{ij}$. From this matrix, we calculated the mean and the standard deviation on $\mathscr{V}$ as a function of the absolute height difference $|h_1-h_2|$ by tracing along the anti-diagonal lines. The shaded regions cover one standard deviation on $\mathscr{V}$. (b) Same as in (a), but the visibilities are shown for a fixed thickness of the waveguide ($220\,\textrm{nm}$) and as a function of the difference on the average waveguide width between the sources. (c) Values of the delays $\tau_1$ (black, relative to source $1$) and $\tau_2$ (green, relative to source $2$) which maximize the visibility in panel (b). \label{Fig_4}}
\end{figure}
We numerically investigated the maximum visibility of two photon interference that can be obtained for increasing amounts of fabrication error. We focused on two key experiments, which are respectively based on the RHOM and on the HHOM effect. In RHOM, the two-photon states $\ket{\textrm{II}}_{1(2)}=\int\Phi_{1(2)}(\omega_s,\omega_i)a_s^{\dagger}(\omega_s)a_i^{\dagger}(\omega_i)\ket{0}$ generated by source $1$ and $2$ are sent at the input ports $a$ and $b$ of a balanced beamsplitter, in the coherent superposition $\ket{\Psi}=\frac{1}{\sqrt{2}} \left ( \ket{\textrm{II}}_{1,a} +e^{i\theta}\ket{\textrm{II}}_{2,b}\right )$. Coincidences are monitored between the output ports as a function of $\theta$. It can be demonstrated that the visibility $\mathscr{V}_{\textrm{RHOM}}$ of the two-photon fringe coincides with the indistinguishability \cite{paesani2020near}, i.e.:
\begin{equation}
\mathscr{V}_{\textrm{RHOM}} = \left | \int \Phi_1(\omega_s,\omega_i)\Phi_2^*(\omega_s,\omega_i) d\omega_s d\omega_i \right |^2.
\label{eq:10}
\end{equation} 
In the case of HHOM, in each source we use one photon of the pair, say the Idler, to herald its partner. Among the heralded Signals, one is delayed with respect to the other, after that the two are interfered at the input ports of a $50/50$ beasmplitter. A dip in the coincidences between the photons emerging at the output ports is observed at zero delay, with visibility \cite{koefoed2017spectrally}:
\begin{equation}
    \mathscr{V}_{\textrm{HHOM}} = \left |\int dT_s dT_s' \left (\Phi_1(T_s,T_i)\Phi_1^*(T_s',T_i)dT_i \right )\left (\Phi_2(T_s',T_i')\Phi_2^*(T_s,T_i')dT_i' \right ) \right |
    \label{eq:11}.
\end{equation}
This quantity depends on both the indistinguishability and the purity of the heralded photons. Figure \ref{Fig_4}(a) shows the maximum values of $\mathscr{V}_{\textrm{RHOM}}$ and $\mathscr{V}_{\textrm{HHOM}}$ which can be achieved after optimization of $\tau_1$ and $\tau_2$. The quantities are shown as a function of the height difference $|h_1-h_2|$ of the two waveguides sources, which are assumed to have the same average width $\langle w \rangle = 2.25\,\mu\textrm{m}$ and tapering $\Delta w=0.25\,\mu\textrm{m}$. In the optimization procedure, the temporal distinguishability is erased in two steps. First, we compute the mean arrival times $(\tau_{s1(2)},\tau_{i1(2)})$ of each photon from the JTAs $\Phi_{1(2)}$. Second, the JTA of source $2$ is shifted in time as $\Phi_2(T_s,T_i)\rightarrow \Phi(T_s+(\tau_{2s}-\tau_ {1s}),T_i+(\tau_{2i}-\tau_{1i}))$ to mimic the presence of a delay stage on the path of each photon. Using the JTA of source $1$ and the delayed JTA of source $2$, the visibilities are computed according to Eq.(\ref{eq:10}-\ref{eq:11}). From Fig.\ref{Fig_4}(a), we see that $\mathscr{V}_{\textrm{RHOM}}$ is higher than $0.95$ for $|h_2-h_1|\leq4.3\,\textrm{nm}$, while in the same range $\mathscr{V}_{\textrm{HHOM}}\ge 0.9$, which is only $3\%$ less than its value at $|h_2-h_1|=0$ ($\mathscr{V}_{\textrm{HHOM}}=0.93$). On the contrary, both $\mathscr{V}_{\textrm{RHOM}}$ and $\mathscr{V}_{\textrm{HHOM}}$ rapidly decrease to zero if the delays are not optimized. As shown in the inset of Fig.\ref{Fig_4}(a), we have that $1\,\textrm{nm}$ of error in the waveguide height is sufficient to drop $\mathscr{V}_{\textrm{RHOM}}$ to $0.89$ and $\mathscr{V}_{\textrm{HHOM}}$ to $0.81$, while their values are almost unaffected ($\mathscr{V}_{\textrm{RHOM}}>0.995$, $(\mathscr{V}_{\textrm{HHOM}}(h_1=h_2)-\mathscr{V}_{\textrm{HHOM}}(\Delta h=1\,\textrm{nm}))/\mathscr{V}_{\textrm{HHOM}}(h_1=h_2)<0.35\%$) in the optimized case. This range is especially relevant for sources which lie on the same die, for which the thickness uniformity of the silicon device layer is sub-nm \cite{siew2021review}. When the fabrication error is considered on the average waveguide width $\langle w \rangle$ (assuming the same height for both sources), a similar result is found. This is shown in Fig.\ref{Fig_4}(b) for $\Delta w=0.1\,\mu\textrm{m}$. As already discussed in Section \ref{sec:principle}, the phase matching wavelengths are less affected by small variations in the waveguide width, reason why for this configuration we choose a smaller tapering amplitude. As an example, for $|\langle w_1 \rangle - \langle w_2 \rangle| \leq60\,\textrm{nm}$ , we have that $\mathscr{V}_{\textrm{RHOM}}\geq 0.995$ and $\mathscr{V}_{\textrm{RHOM}}\geq 0.98$. Without delay optimization, for $|\langle w_1 \rangle - \langle w_2 \rangle| = 60\, \textrm{nm}$ their value drop to $\mathscr{V}_{\textrm{RHOM}}= 0.92$ and $\mathscr{V}_{\textrm{RHOM}}=0.88$. The inset in Fig.\ref{Fig_4}(b) shows that for errors in the waveguide width below $15\,\textrm{nm}$, which is a meaningfull range for sources lying on the same die \cite{siew2021review,selvaraja2009subnanometer}, the optimized values of $\mathscr{V}_{\textrm{RHOM}}$ and $\mathscr{V}_{\textrm{HHOM}}$ are almost equal to the case of identical waveguides. In Fig.\ref{Fig_4}(c) we report the values of $\tau_1$ and $\tau_2$ which maximize the fringe visibility as a function of the error on the average waveguide width. As the latter increases, $\tau_1$ and $\tau_2$ show opposite trends. From the trivial case $\tau_1=\tau_2=\frac{\tau_{\textrm{max}}}{2}$, which occurs at $\langle w_1 \rangle = \langle w_2 \rangle$, we have that by increasing the difference in the waveguide width, $\tau_1$ monotonically decreases while $\tau_2$ increases (we arbitrarily choose $\langle w_1 \rangle \le \langle w_2 \rangle $ to fix the sign of $\Delta\lambda_s$, the behaviour will be inverted in the opposite case). To intuitively understand this trend, suppose that due to an error on the waveguide width, sources $1$ and $2$ emit pairs with a wavelength difference $\Delta\lambda_{s(i)}=\lambda_{s(i)1}-\lambda_{s(i)2}$. We could recover the spectral indistinguishability by acting exclusively on the delay of source $1$, i.e., $\tau_1\rightarrow \frac{\tau_{\textrm{max}}}{2}\pm \Delta\tau$, where the choice of the sign depends on the one of $\Delta\lambda_s$ (equivalently $\Delta\lambda_i$). However, Fig.\ref{Fig_2}(b) indicates that at both large and small delays, the shape of the JSA is asymmetric, and the  purity of the heralded single photon states decreases with respect to $\tau=\frac{\tau_{\textrm{max}}}{2}$. It is then more convenient to modify the delay of both sources, choosing $\tau_1 \sim \frac{\tau_{\textrm{max}}}{2}\pm\frac{\Delta\tau}{2}$ and $\tau_2 \sim \frac{\tau_{\textrm{max}}}{2} \mp \frac{\Delta\tau}{2}$, rather than imparting the whole delay $\Delta\tau$ on source $1$. In this way, the JSA of both sources will have less distortions. \\
\noindent  We evaluated that in order to compensate for silicon device thickness inhomogeneties $\leq1\,\textrm{nm}$ the delay must be tunable in the range $[0.35,0.65]\tau_{\textrm{max}}$, which corresponds to $[1.71,3.18]\,\textrm{ps}$. With reference to the device sketched in Fig.\ref{Fig_1}(b), this could be achieved by placing a delay line in the lower arm after the input beamsplitter, which is reconfigurable in the range $\Delta\tau = \tau_0\pm\frac{\Delta T}{2}$, where $\tau_0$ is a bias delay and $\Delta T = (3.18-1.71) = 1.47\,\textrm{ps}$. When the delay line is set into its rest state ($\Delta\tau=\tau_0$), the length difference between the lower and the upper arm after the input beamsplitter must be $\Delta L = v_{p1}(\frac{\tau_{\textrm{max}}}{2}-\tau_0)$. Among the different devices which can physically implement the delay line, a good candidate is the one based on cascaded asymmetric Mach Zendher Interferometers (aMZI) reported in \cite{waqas2018cascaded}. This device is attractive since it can be easily reconfigured using thermo optic phase shifters, it is built using standard and robust optical components, and has a broadband spectral response. While an in-depth discussion lies out of the scope of this work, we only comment on the feasibility of the method. Following the results found in \cite{waqas2018cascaded}, the maximum delay $\Delta T$ is linked to the FSR of the aMZI as $\textrm{FSR}_{\lambda}=\frac{2\lambda^2}{c\Delta T}$. Using $\Delta T = 1.47\,\textrm{ps}$, we have that $\textrm{FSR}_{\lambda}\sim 11\,\textrm{nm}$, and the minimum $3\,\textrm{dB}$-bandwidth of the device transmittance is $\sim 1.27\times\frac{\textrm{FSR}_{\lambda}}{2}\sim 7\,\textrm{nm}$ \cite{waqas2018cascaded}. This should be sufficiently large to transmit the pump, the Signal and the Idler photons without significant distortions of their temporal wavepackets.  

\section{Influence of SPM and XPM on the source tunability}
\begin{figure}[t!]
\centering\includegraphics[scale = 0.55]{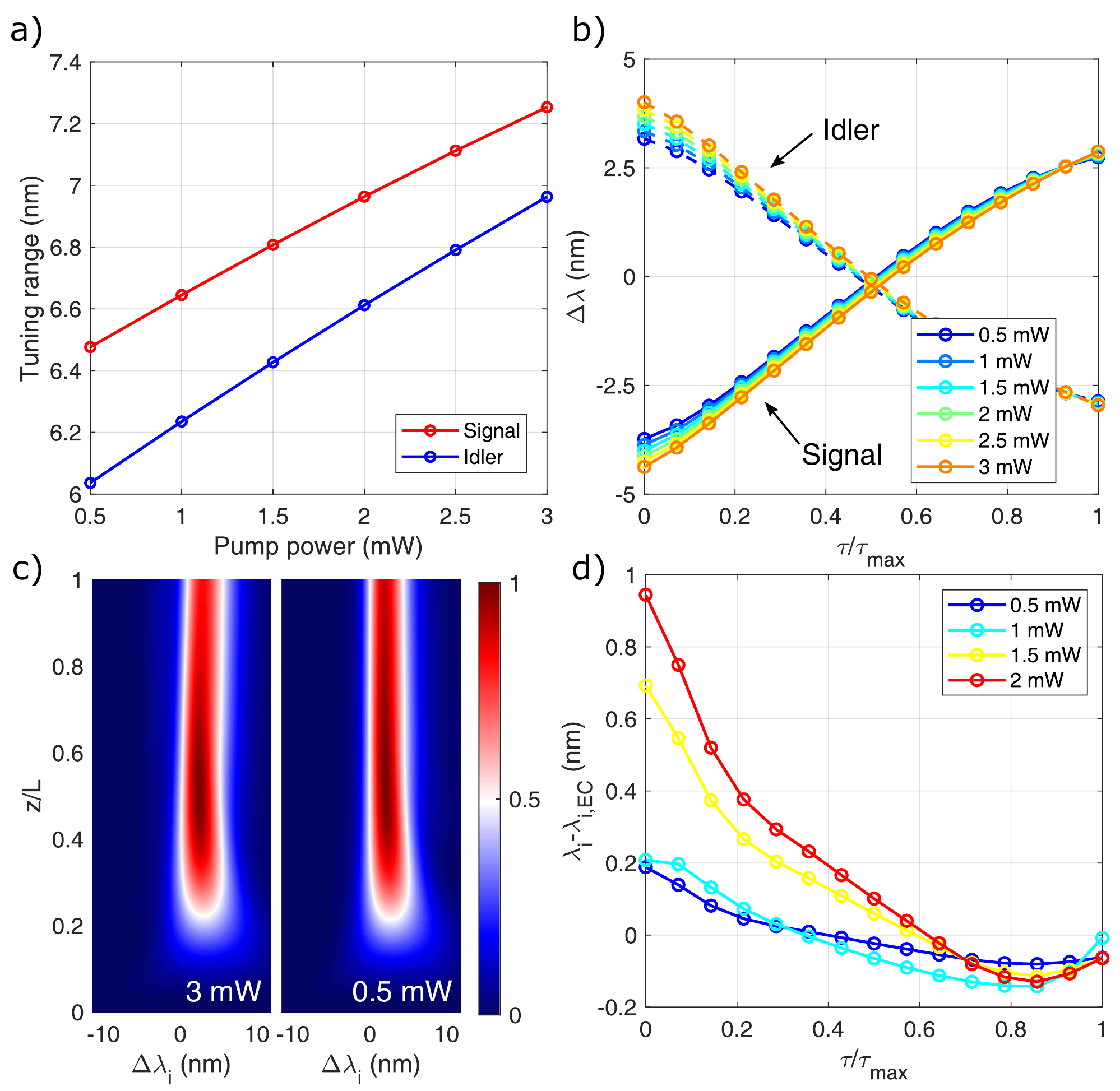}
\caption{(a) Maximum tuning range of the Signal and of the Idler wavelength as a function of the input pump power. (b) Shift of the Signal and Idler wavelength as a function of $\tau$ for different values of the input pump power. In both panels (a) and (b), the tapering amplitude is fixed to $\Delta w=0.25\,\mu\textrm{m}$. (c) Spectrally-resolved cumulative generation probability of the Idler photon as a function of the position along the waveguide. This quantity has been normalized to its maximum for clarity, so it has not to be interpreted as a true probability. The two plots refer to an input power of $3\,\textrm{mW}$ (left) and $0.5\,\textrm{mW}$ (right), while $\tau=0.15\tau_{\textrm{max}}$. (d) Deviation of the average Idler wavelength $\lambda_i$ with respect to the one predicted by energy conservation $\lambda_{i,\textrm{EC}}$. This quantity is shown as a function of $\tau$ and for different values of the input power. \label{Fig_5}}
\end{figure}
\noindent In our scheme, we use sub-ps pulses of high peak power ($\sim 10\,\textrm{W}$ for $1\,\textrm{mW}$ of average power) to compensate the large effective area of the FWM interaction. It is well known that SPM and XPM, triggered by the high power intensities, influence the shape of the JSA \cite{sinclair2016effect,bell2015effects,vernon2015strongly}. In our case, the accumulated SPM of the pumps and their XPM on the Signal/Idler photons both depend  on $\tau$, because the delay determines the position along the waveguide where the two pump pulses overlap and the pair is generated. When $\tau\sim 0$, the SPM accumulated by the pumps is minimum, but the XPM induced on the photons is maximum. The opposite holds when $\tau\sim\tau_{\textrm{max}}$. We numerically simulated these regimes, focusing in particular on how the pump power influences the maximum tuning range $\Delta\lambda^{\textrm{max}}_{s(i)}$. In Fig.\ref{Fig_5}(a) we plot this quantity as a function of the average pump power and a tapering amplitude of $\Delta w =0.25\,\mu\textrm{m}$. The tuning range increases with the pump power, with the Idler photon being slightly more sensitive ($\sim 0.33\frac{\textrm{nm}}{\textrm{mW}}$) than the Signal ($\sim 0.26\frac{\textrm{nm}}{\textrm{mW}}$) to power variations. Figure \ref{Fig_5}(b) shows $\Delta\lambda_{s(i)}$ as a function of $\frac{\tau}{\tau_{\textrm{max}}}$ for different input powers. Nonlinear effects alter the wavelengths of the Signal and the Idler especially at small delays, which suggests that they originate from XPM. As the pump power increases, the Signal blue shifts from the low power condition, while the Idler red shifts. To better understand the origin of this phenomenon, we plot in Fig.\ref{Fig_5}(c) the spectrally-resolved  cumulative probability to generate the Idler along the waveguide, which is obtained by marginalizing $\Phi(\omega_s,\omega_i,z)$ over $\omega_s$. This is shown for $\frac{\tau}{\tau_{\textrm{max}}}=0.15$ in both the low power $(0.5\,\textrm{mW})$ and the high power regime $(3\,\textrm{mW})$. It is evident that, in both cases, Idlers are generated at approximately the same wavelength. We then observe  a red shift and a spectral broadening of the Idler spectra only at high power. This is a clear signature that XPM and SPM are not affecting the phase matching condition, but rather that the spectral shift arises from XPM after that the pair is generated. This phenomenon, called XPM induced asymmetric spectral broadeding, is well known to occur in optical fibers in presence of a temporal walk-off between an intense pump and a weak probe beam \cite{agrawal2000nonlinear}. Since $\Delta\lambda_{s(i)}$ are modified by XPM after that the pair is generated, they do not obey the energy conservation relation $\Delta\lambda_{i,\textrm{EC}}=-\left ( \frac{\lambda_{i0}}{\lambda_{s0}} \right )^2\Delta\lambda_{s,\textrm{EC}}$. This implies that any spectral distinguishability arising from XPM can not be recovered by changing the delay $\tau$. In Fig.\ref{Fig_5}(d) we plot the discrepancy of the Idler wavelength $\lambda_i$ from the one $\lambda_{i,\textrm{EC}}$ expected by energy conservation. To determine $\lambda_{i,\textrm{EC}}$, the pump wavelength is fixed and we use the average wavelength of the Signal extracted from the JSA. At low power, the deviation is zero at $\tau=\frac{\tau_{\textrm{max}}}{2}$, while the small discrepancies at $\tau \rightarrow 0$ and $\tau \rightarrow \tau_{\textrm{max}}$ have exclusively to be attributed to the asymmetric marginal spectra of the Idler which arise from border effects (see Fig.\ref{Fig_2}(b)). Up to $1\,\textrm{mW}$, nonlinear effects still have a limited impact, with $|\lambda_i-\lambda_{i,\textrm{EC}}|\le0.1\,\textrm{nm}$. At $2\,\textrm{mW}$, deviations from energy conservation can be as high as $1\,\textrm{nm}$ at $\tau=0$. However, as shown in Fig.\ref{Fig_4}(c), the delays which are used to correct the fabrication errors lie in the range $\tau\in(0.25,0.7)$, and within this interval  $|\lambda_i-\lambda_{i,\textrm{EC}}|\le0.3\,\textrm{nm}$, which is less than $10\%$ of the spectral linewidth of each photon. We then conclude that, up to $2\,\textrm{mW}$, XPM and SPM effects do not severely compromise the spectral indistinguishability. It is worth to note that the source is conceived to work in the low (e.g., $\xi<0.1$) squeezing regime to limit multiphoton contamination in the heralded photon states \cite{bonneau2015effect}. Therefore, it is very unlikely that we will use input powers higher than $1\,\textrm{mW}$, since this level already corresponds to $\xi=0.1$ (see Fig.\ref{Fig_2}(d)). As a comparison, we have that $\xi=0.25$ at $2\,\textrm{mW}$ of input power. 
\section{Conclusions}
\label{sec:Conclusions}
\noindent We proposed a scheme to generate highly pure and spectrally tunable photon pairs using delayed-pump Intermodal Four Wave Mixing. The high purity is inherited from the engineering of the phase matching relation and from the adiabatic switching of the nonlinear interaction. The tunability of the emission wavelength is added by tapering the width of the waveguide, and by changing the delay between the pump pulses. We demonstrate that the tunability range can be extended by increasing the tapering amplitude, with only a modest reduction in the purity of the heralded single photon states and with almost no impact on the pair generation probability. We show that, by optimizing the pump delay, we can drastically reduce the distinguishability among independent sources which arise from fabrication errors. Under realistic fabrication tolerances, an indistinguishability level $>95\%$ can be guaranteed up to a difference in the waveguide height of $4\,\textrm{nm}$, and for errors in the waveguide width larger than $100$ nm. Under these circumstances, we predicted a degradation of the HHOM visibility of less than  $3\%$ of its value compared to the case of two identical sources. In comparison, the visibility and the indistinguishability will be both below $20\%$ without delay optimization. We also show that, in the regime of low pair generation, XPM and SPM effects are not affecting the device performance. The proposed device can be built using standard integrated optical components provided by commercial photonic design kits and could be reconfigured using thermo optical phase shifters. Its implementation can mitigate indistinguishability issues either in large scale quantum photonic circuits encompassing arrays of sources, or in distant devices for quantum communication which are manufactured on different chips. 

\section*{Appendix A: Expressions for the linear and nonlinear momentum}
\noindent The momentum flux $M$ governing the spatial evolution along the waveguide length of each operator is defined as \cite{huttner1990quantum}:
\begin{equation}
    M(z) = \int (\mathbf{D} \times \mathbf{B})\cdot \hat{z} dxdy \sim \frac{1}{c}\int  D_y^{-}(x,y,z,t)E_y^{+}(x,y,z,t)  dx dy dt + \textrm{h.c.}, \label{eq:AA1}
\end{equation}
where $D^{-}$ and $E^{+}$ denote respectively the negative and the positive frequency part of the displacement and the electric field operator. The first involves only photon creation operators, while the second only annihilation operators (see the field expansion in Eq.(\ref{eq:2})). In Eq.(\ref{eq:AA1}), we assumed that the $\mathbf{E}$ field is entirely polarized along the $y$ direction (TM modes), and that $\mathbf{B}$ can be expressed as $\mathbf{B}=\frac{1}{c}(E_y,0,0)$ (plane wave approximation). Then, one writes $D_y=\epsilon_0 n^2 E_y+P_y^{\textrm{NL}}$, where $n$ is the material refractive index and $P_y^{\textrm{NL}}$ is the nonlinear polarization, which in our case consists only in the term $P_y^{\textrm{NL}}=\epsilon_0\chi^{(3)}_{yyyy}E_y^3$, where  $\chi^{(3)}_{yyyy}$ is the isotropic contribution to the third order nonlinear susceptibility. In the next steps, one finds suitable expressions for $\mathbf{E}(x,y,z,t)$, as the ones in Eq.(\ref{eq:2}), insert them into Eq.(\ref{eq:AA1}), and separates the linear terms from the ones generated by the nonlinear polarization. This standard procedure can be found, e.g., in \cite{quesada2021beyond}, hence we will only report the final result. The linear momentum $M_L$ is given by \cite{quesada2021beyond,sinclair2016effect}:
\begin{equation}
    M_L = \int \hbar \beta(\omega) a^{\dagger}(\omega,z)a(\omega,z) d\omega.  \label{eq:AA2}
\end{equation}
Since the pump, the Signal and the Idler fields are narrowband and centered into three non-overlapping frequency ranges, the integral in Eq.(\ref{eq:AA2}) can be split into $M_L = M^{p}_{L}+M^{s}_{L}+M^{i}_{L}$. Each term has the same form of Eq.(\ref{eq:AA2}), but with the integral restricted to the frequency range of the corresponding beam. Within these intervals, one can define the slowly varying operators (see Eq.(\ref{eq:2})) $\Tilde{a}_q(z,\omega_q)=a(\omega_q+\bar{\omega}_q,z)e^{-i\bar{\beta}_{q}z}$, where $q=\{p,s,i\}$. By Taylor expanding the wavevector $\beta$ up to the second order in the frequency detuning $(\omega-\bar{\omega})$ in Eq.(\ref{eq:AA2}), we have:
\begin{equation}
    M_L = \sum_{q=\{p,s,i\}} \hbar \int (\bar{\beta}_{q}+v_q(\omega_q-\bar{\omega}_{q})+\frac{\beta^{(2)}_q}{2} (\omega_q-\bar{\omega}_{q})^2)\Tilde{a}^{\dagger}_q(\omega_q-\bar{\omega}_q,z)\Tilde{a}_q(\omega_q-\bar{\omega}_q,z) d\omega_q, \label{eq:AA3}
\end{equation}
where $\beta_q^{(2)}=\frac{d^2\beta_q}{d\omega^2}$. The SPM, XPM and FWM terms are  more easily expressed in the time domain, where they have the following form \cite{sinclair2016effect}:
\begin{equation}
\label{eq:AA4}
\begin{aligned}
M_{\textrm{SPM}} = & \frac{1}{2}\hbar \sum_{q=\{ p1, p2\}}\gamma_{qqqq}\int |\Tilde{A}_q(z,t)|^2 \Tilde{A}_q(z,t) dt, \\
M_{\textrm{XPM}} = & 2\hbar \sum_{q = \{ p1,p2\}}\sum_{r = \{ s,i\}} \gamma_{qqrr}|\Tilde{A}_q(z,t)|^2 \Tilde{A}_r(z,t) dt, \\
M_{\textrm{FWM}} = & \hbar \gamma_{p1p2si}\int \Tilde{A}_{p1}(z,t)\Tilde{A}_{p2}(z,t)\Tilde{a}_s^{\dagger}(z,t)\Tilde{a}_i^{\dagger}(z,t) dt. 
\end{aligned}
\end{equation}
The definitions in Eq.(\ref{eq:AA4}) make use of the nonlinear parameter $\gamma_{\textrm{ijkl}}=\frac{\omega n_2}{cA_{\textrm{ijkl}}}$, where $n_2$ is the nonlinear refractive index of silicon and $A_{\textrm{ijkl}}$ is the nonlinear effective area, defined as:
\begin{equation}
    A_{\textrm{ijkl}} = \frac{\prod_{q = \{ i,j,k,l\}} \left ( \int |F_q(x,y)|^2 dx dy \right )^{\frac{1}{2}} }{\int_{\textrm{wg}} F_i(x,y)F_j(x,y) F_k(x,y)^{*} F_l(x,y)^{*} dx dy}. \label{eq:AA5}
\end{equation}
The values of the parameters introduced so far and the ones appearing in Eqs.(\ref{eq:3}-\ref{eq:7}) are calculated using the commercial Lumerical MODE package \cite{Lumerical}, and are listed in Table \ref{Tab1}. Losses are taken from the measured values in \cite{paesani2020near}.
\begin{table}[h!]
\begin{center}
\begin{tabular}{ |c|c| } 
 \hline
 Parameter & Value  \\ 
 \hline
 $\alpha_{p1}$ & $0.4\,\textrm{dB}/\textrm{cm}$ \\ \hline
 $\alpha_{p2}$ & $0.2\,\textrm{dB}/\textrm{cm}$ \\ \hline
 $v_{p1}$ & $75.20\,\mu\textrm{m}/\textrm{ps}$ \\ \hline
 $v_{p2}$ & $73.41\,\mu\textrm{m}/\textrm{ps}$ \\ \hline 
 $v_{s}$ & $75.29\,\mu\textrm{m}/\textrm{ps}$ \\ \hline
 $v_{i}$ & $73.5\,\mu\textrm{m}/\textrm{ps}$ \\ \hline 
 $L_{w_p}$ & $0.25\,\textrm{cm}$ \\ \hline
 $|L_{w_s}|$ & $3.27\,\textrm{cm}$ \\ \hline 
 $|L_{w_i}|$ & $0.26\,\textrm{cm}$ \\ \hline
 $L_{D_{p1}}$ & $4.60\,\textrm{cm}$ \\ \hline
 $L_{D_{p2}}$ & $4.43\,\textrm{cm}$ \\ \hline
 $L_{D_{s}}$ & $4.09\,\textrm{cm}$ \\ \hline
 $L_{D_{i}}$ & $5.18\,\textrm{cm}$ \\ \hline
 $\gamma_{1111}$ & $2.73\,\frac{1}{\textrm{m}\cdot\textrm{W}}$ \\ \hline
 $\gamma_{1122} = \gamma_{2211}$ & $1.77\,\frac{1}{\textrm{m}\cdot\textrm{W}}$ \\ \hline
 $\gamma_{2222}$ & $2.60\,\frac{1}{\textrm{m}\cdot\textrm{W}}$ \\ \hline
 $\gamma_{11ss}$ & $1.52\,\frac{1}{\textrm{m}\cdot\textrm{W}}$ \\ \hline   
 $\gamma_{22ss}$ & $2.25\,\frac{1}{\textrm{m}\cdot\textrm{W}}$ \\ \hline   
 $\gamma_{11ii}$ & $3.12\,\frac{1}{\textrm{m}\cdot\textrm{W}}$ \\ \hline   
 $\gamma_{22ii}$ & $2.01\,\frac{1}{\textrm{m}\cdot\textrm{W}}$ \\ \hline   
 $\gamma_{p1p2si}$ & $1.34\,\frac{1}{\textrm{m}\cdot\textrm{W}}$ \\ \hline   
\end{tabular}
\end{center}
\caption{Values of the parameters used in the simulation of the JSA and JTA through this work. \label{Tab1}}
\end{table}
\section*{Appendix B: Derivation of the propagation equation for the JTA}
\noindent We start from the definition of the JTA given in Section \ref{sec:principle}, that we rewrite here for clarity:
\begin{equation}
    \Phi(T_s,T_i,z) =  \langle \Tilde{a}_s(T_s,z)\Tilde{a}_i(T_i,z)U(z,0)\rangle.  \label{eq:A1} 
\end{equation}
By performing the derivative in $z$ of both members in Eq.(\ref{eq:A1}) we get:
\begin{equation}
    \frac{\partial \Phi}{\partial z} = \langle  \frac{\partial \Tilde{a}_s}{\partial z}\Tilde{a}_i U+\Tilde{a}_s\frac{\partial \Tilde{a_i}}{\partial z} U\rangle+\langle \Tilde{a}_s\Tilde{a}_i\frac{\partial U}{\partial z}\rangle.  \label{eq:A2}
\end{equation}
We now use Eq.(\ref{eq:5}) to express $\frac{\partial \Tilde{a}_{s(i)}}{\partial z}=(L'_{s(i)} + N_{s(i)})\Tilde{a}_{s(i)}$, where  $L'_{s(i)}=L_{s(i)}+i\Delta\beta_{s(i)}(z)$ and the definition of the operators $L_{s(i)}$ and $N_{s(i)}$ are given in Eq.(\ref{eq:7}). Moreover, from Section \ref{sec:principle} we have that $\frac{\partial U}{\partial z} = \frac{i}{\hbar} M_{\textrm{FWM}}$, so as Eq.(\ref{eq:A2}) becomes:
\begin{equation}
    \frac{\partial \Phi}{\partial z} =  \langle[(L'_s+ N_s)\Tilde{a}_s]\Tilde{a}_i U\rangle+\langle \Tilde{a}_s[(L'_i+N_i)\Tilde{a}_i]U\rangle +\frac{i}{\hbar}\langle \Tilde{a}_s\Tilde{a}_i M_{\textrm{FWM}}\rangle. \label{eq:A3}
\end{equation}
The first two terms on the right hand side have exactly the same form, so we will only treat the case of the Signal and apply the same result to the Idler. Writing $\Tilde{a}_s(T_s,z)$ using its Fourier Transform:
\begin{equation}
    \Tilde{a}_s(T_s,z) = \int \Tilde{a}_s(\omega_s',z)e^{-iT_s\omega_s'}d\omega_s', \label{eq:A4}
\end{equation}
the application of $L'_s$ replaces $\Tilde{a}_s$ by $\Tilde{a}_s\rightarrow f(\omega_s')\Tilde{a}_s = (-\alpha_s/2+i\Delta\beta_s \omega_s'/L_{w_s}+i\omega_s'^2/(2L_{D_s}))\Tilde{a}_s$. By writing also $\Tilde{a_i}$ using its Fourier Transform, we have:
\begin{equation}
    \langle[(L'_s+N_s)\Tilde{a}_s]\Tilde{a}_i U\rangle = \int f(\omega_s')e^{-iT_s\omega_s'}e^{-iT_i\omega_i'}\langle \Tilde{a}_s(\omega_s',z)\Tilde{a}_i(\omega_i',z) U \rangle d\omega_s' d\omega_i'. \label{eq:A5}
\end{equation}
The expectation value on the right hand side is the definition of the JSA $\Tilde{\Phi}(\omega_s',\omega_i',z)$. Following an identical procedure for the Idler, we have that:
\begin{equation}
    \langle [L'_s\Tilde{a_s}]\Tilde{a_i}U+\Tilde{a}_s [L'_i\Tilde{a_i}]U\rangle = \int [f(\omega_s')+f(\omega_i')]e^{-iT_s\omega_s'}e^{-iT_i\omega_i'}\Phi(\omega_s',\omega_i',z)d\omega_s'd\omega_i', \label{eq:A6}
\end{equation}
which is equivalent to apply the operator $L'_{s(i)}$ to the JTA. In the time domain, the operator $N_{s(i)}$ is simply a multiplicative factor, so we have that $\langle N_{s(i) }\Tilde{a}_s\Tilde{a}_i U \rangle= N_{s(i)} \langle \Tilde{a}_s\Tilde{a}_i U \rangle = N_{s(i)}(T_{s(i)})\Phi(T_s,T_i,z)$. 
By combining this result with the expression in Eq.(\ref{eq:A6}), the first two terms on the right hand side of Eq.(\ref{eq:A3}) becomes $\langle[(L'_s+N_s)\Tilde{a}_s]\Tilde{a}_i U\rangle+\langle \Tilde{a}_s[(L'_i+N_i)\Tilde{a}_i]U\rangle=(L'_s+L'_i+N_s+N_i)\Phi(T_s,T_i,z)$. We now work out the driving term $\langle \Tilde{a}_s\Tilde{a}_i M_{\textrm{FWM}}\rangle$ in the frequency domain. By using Eq.(\ref{eq:A4}), we have:
\begin{equation}
\label{eq:A7}  
\begin{aligned}
\frac{i}{\hbar}\langle \Tilde{a}_s\Tilde{a}_i M_{\textrm{FWM}}\rangle = & i\gamma_{p1p2si}\int A_{p1}(t,z)A_{p2}(t,z)\langle \Tilde{a}_s(\omega_s,z)\Tilde{a}_s(\omega_i,z)\Tilde{a}^{\dagger}_s(\omega_s',z)\Tilde{a}^{\dagger}_i(\omega_i',z)\rangle \times, \\ 
{} & e^{-i(T_s\omega_s+T_i\omega_i)}e^{iT(\omega_s'+\omega_i')}dt d\omega_s d\omega_s' d\omega_i d\omega_i'. 
\end{aligned}
\end{equation}
Using the equal space commutation relations $[a(\omega,z),a^{\dagger}(\omega',z)]=\delta(\omega-\omega')$, the expectation value in Eq.(\ref{eq:A7}) gives $\delta(\omega_s-\omega_s')\delta(\omega_i-\omega_i')$. After integration in $\omega_s$ and $\omega_i$, we have $\omega_s=\omega_s'$ and $\omega_i = \omega_i'$. Then, in the frequency domain, the product of the pump envelopes is $A_{p1}(t,z)A_{p2}(t,z)=\int \mathscr{A}_{p1}(x',z)\mathscr{A}_{p2}(x-x',z)e^{ixt}dx dx'$, which inserted into Eq.(\ref{eq:A7}) gives:
\begin{equation}
    \frac{i}{\hbar}\langle \Tilde{a}_s\Tilde{a}_i M_{\textrm{FWM}}\rangle = i\gamma_{p1p2si}\int \mathscr{A}_{p1}(x,z)\mathscr{A}_{p2}(x-x',z)e^{it(\omega_s'+\omega_i'-x)}e^{-i(T_s\omega_s'+T_i\omega_i')}d\omega_s' d\omega_i' dt dx dx'.
    \label{eq:A8}
\end{equation}
Integration over $t$ gives $\delta(\omega_s'+\omega_i'-x)$, and a subsequent integration over $x$ sets $x=\omega_s'+\omega_i'$. Expression Eq.(\ref{eq:A8}) reduces to:
\begin{equation}
\frac{i}{\hbar}\langle \Tilde{a}_s\Tilde{a}_i M_{\textrm{FWM}}\rangle = e^{i\int_0^z (\Delta\beta_{p1}(z')+\Delta\beta_{p2}(z'))dz'}\int G(\omega_s',\omega_i',z)e^{-i(T_s\omega_s'+T_i\omega_i')}d\omega_s' d\omega_i', \label{eq:A9}
\end{equation}
where $G=i\int \bar{\mathscr{A}}_{p1}(x,z)\bar{\mathscr{A}}_{p2}(\omega_s'+\omega_i'-x) dx$ and we wrote the pump as $\mathscr{A}_{p1(2)}(x,z)=\bar{\mathscr{A}}_{p1(2)}(x,z)\exp{\left ( i\int_0^z \Delta\beta_{p1(2)}(z')dz' \right )}$ to factor out the effect of tapering. As a final step, in order to recover Eq.(\ref{eq:7}) of the main text, we write $\Phi(T_s,T_i,z)=\Tilde{\Phi}(T_s,T_i,z)e^{i\Theta_{si}(z)}$, where $\Theta_{si}$ is defined in Eq.(\ref{eq:7}), and we use the fact that $\frac{\partial \Theta_{si}}{\partial z}=(\Delta\beta_s+\Delta\beta_i)$ cancels the $\Delta\beta$ terms in the operators $L'_{s(i)}$.  

\section*{Appendix C: Analytic expression for the cumulative pair generation probability}
\noindent Here we derive an analytic expression for the evolution of the cumulative pair generation probability along the waveguide, in the approximation of zero loss, SPM, XPM and group velocity dispersion. In this scenario, Eq.(\ref{eq:8}) reduces to $\xi(z)=2\int_0^z \mathscr{R}\left [ S\Phi^*\right]$. By using the expression of $S$ in Eq.(\ref{eq:7}), we get:
\begin{equation}
    \xi(z) = 2\int_0^z \mathscr{R} \left [ A_{p1}(z',t)A_{p2}(z',t)\Phi^*\exp{(it(\omega_s'+\omega_i')-T_s\omega_s'-T_i\omega_i')}d\omega_s'd\omega_i'dT_sdT_i dz' dt \right ], \label{eq:B1}
\end{equation}
where we have used $\int \bar{\mathscr{A}}_{p1}(x,z)\bar{\mathscr{A}}_{p2}(\omega_s'+\omega_i'-x,z)dx = \int A_{p1}(t,z)A_{p2}(t,z)e^{it(\omega_s'+\omega_i')}dt$. The phase mismatch $e^{i\Theta}$ term in Eq.(\ref{eq:7}) has been assumed to be $\sim 1$ in the region where the product of the pump envelopes is $\neq 0$. We now integrate the exponential in the two variables $\omega_s'$ and $\omega_i'$ to obtain the product of the Dirac delta $\delta(t-T_s)\delta(t-T_i)$, which is not vanishing only when $T_s=T_i=t$. Hence Eq.(\ref{eq:B1}) reduces to:
\begin{equation}
    \xi(z) = 2\int_0^z\mathscr{R} \left [ A_{p1}(t,z')A_{p2}(t,z')\Phi(T_s=t,T_i=t,z')^*\right ]dz'dt. \label{eq:B2}
\end{equation}
In absence of group velocity dispersion, XPM and SPM, we have that $\Phi\propto A_{p1}(t_c,z_c)A_{p_2}(t_c,z_c)$, in which $t_c$ and $z_c$ define the collision time and the collision coordinate of the pair generation event \cite{bell2015effects,koefoed2017spectrally,koefoed2019complete}. These are given by:
\begin{equation}
    z_c = L-\frac{T_0(T_s-T_i)}{v_s^{-1}-v_i^{-1}}, \qquad t_c = \frac{v_s^{-1}T_i-v_i^{-1}T_s}{v_s^{-1}-v_i^{-1}}. \label{eq:B3}
\end{equation}
From these definitions, we see that when $T_s=T_i=t$, the collision coordinates are $z_c=L$ and $t_c=t$. From a physical point of view, this reflects the fact that pairs are generated at the same time but they possess different group velocities. Hence, in order to be detected at the same time, the collision coordinate should coincide with the end of the waveguide. Using these results, Eq.(\ref{eq:B2}) becomes:
\begin{equation}
    \xi(z)\propto \int_0^z |A_{p1}(z',t)|^2|A_{p2}(z',t)|^2 dz'dt. \label{eq:B4}
\end{equation}
We can evaluate the time integral in Eq.(\ref{eq:B4}) by moving off from the pump reference frame, in which the pump envelopes have expression $A_{p1}=\bar{A}_{p1}e^{-\frac{z-v_1 (t-\tau)}{2\sigma_1^2}}$ and $A_{p2}=\bar{A}_{p2}e^{-\frac{z-v_2 t}{2\sigma_2^2}}$, where $\sigma_{1(2)}$ are given by $\sigma_{1(2)}=\frac{v_{1(2)}T_0}{2\sqrt{\ln2}}$ and $|\bar{A}_{p1(2)}|^2$ is the peak powerof the pulses. After the time integral, the result is a gaussian function centered at $z=L_{\textrm{match}}=\tau \left ( \frac{1}{v_2}-\frac{1}{v_1}\right )=\frac{\tau}{\tau_{\textrm{max}}}L$ and with standard deviation $\sigma_z$:
\begin{equation}
    \sigma_z = \frac{\sqrt{v_2^2\sigma_1^2+v_1^2\sigma_2^2}}{v_1^2-v_2^2} = \frac{L_{w_p}}{2\sqrt{\ln 2}},\label{eq:B5}
\end{equation}
which is the result of the main text. Therefore:
\begin{equation}
    \xi(z)\propto \int_0^z e^{-\frac{(z'-L_{\textrm{match}})^2}{2\sigma_z^2}}dz' \propto \textrm{erf}\left (\frac{z-L_{\textrm{match}}}{\sqrt{2}\sigma_z} \right ). \label{eq:B6}
\end{equation}

\section*{Funding}
This work has been funded by the H2020 European project EPIQUS (EC 899368).



\begin{thebibliography}{10}
\newcommand{\enquote}[1]{``#1''}

\bibitem{caspani2017integrated}
L.~Caspani, C.~Xiong, B.~J. Eggleton, D.~Bajoni, M.~Liscidini, M.~Galli,
  R.~Morandotti, and D.~J. Moss, \enquote{Integrated sources of photon quantum
  states based on nonlinear optics,} {\protect\JournalTitle{Light: Science \&
  Applications}} \textbf{6}, e17100--e17100 (2017).

\bibitem{wang2020integrated}
J.~Wang, F.~Sciarrino, A.~Laing, and M.~G. Thompson, \enquote{Integrated
  photonic quantum technologies,} {\protect\JournalTitle{Nature Photonics}}
  \textbf{14}, 273--284 (2020).

\bibitem{bonneau2015effect}
D.~Bonneau, G.~J. Mendoza, J.~L. O’Brien, and M.~G. Thompson, \enquote{Effect
  of loss on multiplexed single-photon sources,} {\protect\JournalTitle{New
  Journal of Physics}} \textbf{17}, 043057 (2015).

\bibitem{collins2013integrated}
M.~J. Collins, C.~Xiong, I.~H. Rey, T.~D. Vo, J.~He, S.~Shahnia, C.~Reardon,
  T.~F. Krauss, M.~Steel, A.~S. Clark \emph{et~al.}, \enquote{Integrated
  spatial multiplexing of heralded single-photon sources,}
  {\protect\JournalTitle{Nature communications}} \textbf{4}, 1--7 (2013).

\bibitem{renema2018efficient}
J.~J. Renema, A.~Menssen, W.~R. Clements, G.~Triginer, W.~S. Kolthammer, and
  I.~A. Walmsley, \enquote{Efficient classical algorithm for boson sampling
  with partially distinguishable photons,} {\protect\JournalTitle{Physical
  review letters}} \textbf{120}, 220502 (2018).

\bibitem{shchesnovich2014sufficient}
V.~Shchesnovich, \enquote{Sufficient condition for the mode mismatch of single
  photons for scalability of the boson-sampling computer,}
  {\protect\JournalTitle{Physical Review A}} \textbf{89}, 022333 (2014).

\bibitem{sparrow2017quantum}
C.~Sparrow, \enquote{Quantum interference in universal linear optical devices
  for quantum computation and simulation,} {\protect\JournalTitle{PhD Thesis}}
  (2017).

\bibitem{signorini2020chip}
S.~Signorini and L.~Pavesi, \enquote{On-chip heralded single photon sources,}
  {\protect\JournalTitle{AVS Quantum Science}} \textbf{2}, 041701 (2020).

\bibitem{paesani2019generation}
S.~Paesani, Y.~Ding, R.~Santagati, L.~Chakhmakhchyan, C.~Vigliar, K.~Rottwitt,
  L.~K. Oxenl{\o}we, J.~Wang, M.~G. Thompson, and A.~Laing, \enquote{Generation
  and sampling of quantum states of light in a silicon chip,}
  {\protect\JournalTitle{Nature Physics}} \textbf{15}, 925--929 (2019).

\bibitem{arrazola2021quantum}
J.~Arrazola, V.~Bergholm, K.~Br{\'a}dler, T.~Bromley, M.~Collins, I.~Dhand,
  A.~Fumagalli, T.~Gerrits, A.~Goussev, L.~Helt \emph{et~al.}, \enquote{Quantum
  circuits with many photons on a programmable nanophotonic chip,}
  {\protect\JournalTitle{Nature}} \textbf{591}, 54--60 (2021).

\bibitem{vigliar2021error}
C.~Vigliar, S.~Paesani, Y.~Ding, J.~C. Adcock, J.~Wang, S.~Morley-Short,
  D.~Bacco, L.~K. Oxenl{\o}we, M.~G. Thompson, J.~G. Rarity \emph{et~al.},
  \enquote{Error-protected qubits in a silicon photonic chip,}
  {\protect\JournalTitle{Nature Physics}} \textbf{17}, 1137--1143 (2021).

\bibitem{adcock2018hard}
J.~C. Adcock, S.~Morley-Short, J.~W. Silverstone, and M.~G. Thompson,
  \enquote{Hard limits on the postselectability of optical graph states,}
  {\protect\JournalTitle{Quantum Science and Technology}} \textbf{4}, 015010
  (2018).

\bibitem{adcock2019programmable}
J.~C. Adcock, C.~Vigliar, R.~Santagati, J.~W. Silverstone, and M.~G. Thompson,
  \enquote{Programmable four-photon graph states on a silicon chip,}
  {\protect\JournalTitle{Nature communications}} \textbf{10}, 1--6 (2019).

\bibitem{llewellyn2020chip}
D.~Llewellyn, Y.~Ding, I.~I. Faruque, S.~Paesani, D.~Bacco, R.~Santagati, Y.-J.
  Qian, Y.~Li, Y.-F. Xiao, M.~Huber \emph{et~al.}, \enquote{Chip-to-chip
  quantum teleportation and multi-photon entanglement in silicon,}
  {\protect\JournalTitle{Nature Physics}} \textbf{16}, 148--153 (2020).

\bibitem{graffitti2018independent}
F.~Graffitti, P.~Barrow, M.~Proietti, D.~Kundys, and A.~Fedrizzi,
  \enquote{Independent high-purity photons created in domain-engineered
  crystals,} {\protect\JournalTitle{Optica}} \textbf{5}, 514--517 (2018).

\bibitem{burridge2020high}
B.~M. Burridge, I.~I. Faruque, J.~G. Rarity, and J.~Barreto, \enquote{High
  spectro-temporal purity single-photons from silicon micro-racetrack
  resonators using a dual-pulse configuration,} {\protect\JournalTitle{Optics
  Letters}} \textbf{45}, 4048--4051 (2020).

\bibitem{liu2020high}
Y.~Liu, C.~Wu, X.~Gu, Y.~Kong, X.~Yu, R.~Ge, X.~Cai, X.~Qiang, J.~Wu, X.~Yang
  \emph{et~al.}, \enquote{High-spectral-purity photon generation from a
  dual-interferometer-coupled silicon microring,} {\protect\JournalTitle{Optics
  Letters}} \textbf{45}, 73--76 (2020).

\bibitem{blay2017effects}
D.~R. Blay, M.~Steel, and L.~Helt, \enquote{Effects of filtering on the purity
  of heralded single photons from parametric sources,}
  {\protect\JournalTitle{Physical Review A}} \textbf{96}, 053842 (2017).

\bibitem{siew2021review}
S.~Y. Siew, B.~Li, F.~Gao, H.~Y. Zheng, W.~Zhang, P.~Guo, S.~W. Xie, A.~Song,
  B.~Dong, L.~W. Luo \emph{et~al.}, \enquote{Review of silicon photonics
  technology and platform development,} {\protect\JournalTitle{Journal of
  Lightwave Technology}} \textbf{39}, 4374--4389 (2021).

\bibitem{silverstone2015qubit}
J.~W. Silverstone, R.~Santagati, D.~Bonneau, M.~J. Strain, M.~Sorel, J.~L.
  O’Brien, and M.~G. Thompson, \enquote{Qubit entanglement between
  ring-resonator photon-pair sources on a silicon chip,}
  {\protect\JournalTitle{Nature communications}} \textbf{6}, 1--7 (2015).

\bibitem{meyer2017limits}
E.~Meyer-Scott, N.~Montaut, J.~Tiedau, L.~Sansoni, H.~Herrmann, T.~J. Bartley,
  and C.~Silberhorn, \enquote{Limits on the heralding efficiencies and spectral
  purities of spectrally filtered single photons from photon-pair sources,}
  {\protect\JournalTitle{Physical Review A}} \textbf{95}, 061803 (2017).

\bibitem{kumar2013spectrally}
R.~Kumar, J.~R. Ong, J.~Recchio, K.~Srinivasan, and S.~Mookherjea,
  \enquote{Spectrally multiplexed and tunable-wavelength photon pairs at 1.55
  $\mu$m from a silicon coupled-resonator optical waveguide,}
  {\protect\JournalTitle{Optics letters}} \textbf{38}, 2969--2971 (2013).

\bibitem{jin2013widely}
R.-B. Jin, R.~Shimizu, K.~Wakui, H.~Benichi, and M.~Sasaki, \enquote{Widely
  tunable single photon source with high purity at telecom wavelength,}
  {\protect\JournalTitle{Optics express}} \textbf{21}, 10659--10666 (2013).

\bibitem{zhu2021spectral}
D.~Zhu, C.~Chen, M.~Yu, L.~Shao, Y.~Hu, C.~Xin, M.~Yeh, S.~Ghosh, L.~He,
  C.~Reimer \emph{et~al.}, \enquote{Spectral control of nonclassical light
  using an integrated thin-film lithium niobate modulator,}
  {\protect\JournalTitle{arXiv preprint arXiv:2112.09961}}  (2021).

\bibitem{li2016efficient}
Q.~Li, M.~Davan{\c{c}}o, and K.~Srinivasan, \enquote{Efficient and low-noise
  single-photon-level frequency conversion interfaces using silicon
  nanophotonics,} {\protect\JournalTitle{Nature Photonics}} \textbf{10},
  406--414 (2016).

\bibitem{paesani2020near}
S.~Paesani, M.~Borghi, S.~Signorini, A.~Ma{\"\i}nos, L.~Pavesi, and A.~Laing,
  \enquote{Near-ideal spontaneous photon sources in silicon quantum photonics,}
  {\protect\JournalTitle{Nature communications}} \textbf{11}, 1--6 (2020).

\bibitem{signorini2018intermodal}
S.~Signorini, M.~Mancinelli, M.~Borghi, M.~Bernard, M.~Ghulinyan, G.~Pucker,
  and L.~Pavesi, \enquote{Intermodal four-wave mixing in silicon waveguides,}
  {\protect\JournalTitle{Photonics Research}} \textbf{6}, 805--814 (2018).

\bibitem{signorini2021silicon}
S.~Signorini, M.~Sanna, S.~Piccione, M.~Ghulinyan, P.~Tidemand-Lichtenberg,
  C.~Pedersen, and L.~Pavesi, \enquote{A silicon source of heralded single
  photons at 2 $\mu$ m,} {\protect\JournalTitle{APL Photonics}} \textbf{6},
  126103 (2021).

\bibitem{christ2011probing}
A.~Christ, K.~Laiho, A.~Eckstein, K.~N. Cassemiro, and C.~Silberhorn,
  \enquote{Probing multimode squeezing with correlation functions,}
  {\protect\JournalTitle{New Journal of Physics}} \textbf{13}, 033027 (2011).

\bibitem{koefoed2017effects}
J.~G. Koefoed, J.~B. Christensen, and K.~Rottwitt, \enquote{Effects of
  noninstantaneous nonlinear processes on photon-pair generation by spontaneous
  four-wave mixing,} {\protect\JournalTitle{Physical Review A}} \textbf{95},
  043842 (2017).

\bibitem{agrawal1989temporal}
G.~P. Agrawal, P.~Baldeck, and R.~Alfano, \enquote{Temporal and spectral
  effects of cross-phase modulation on copropagating ultrashort pulses in
  optical fibers,} {\protect\JournalTitle{Physical Review A}} \textbf{40}, 5063
  (1989).

\bibitem{agrawal2000nonlinear}
G.~P. Agrawal, \enquote{Nonlinear fiber optics,} in \emph{Nonlinear Science at
  the Dawn of the 21st Century,}  (Springer, 2000), pp. 195--211.

\bibitem{huttner1990quantum}
B.~Huttner, S.~Serulnik, and Y.~Ben-Aryeh, \enquote{Quantum analysis of light
  propagation in a parametric amplifier,} {\protect\JournalTitle{Physical
  Review A}} \textbf{42}, 5594 (1990).

\bibitem{sinclair2016effect}
G.~F. Sinclair and M.~G. Thompson, \enquote{Effect of self-and cross-phase
  modulation on photon pairs generated by spontaneous four-wave mixing in
  integrated optical waveguides,} {\protect\JournalTitle{Physical Review A}}
  \textbf{94}, 063855 (2016).

\bibitem{bell2015effects}
B.~Bell, A.~McMillan, W.~McCutcheon, and J.~Rarity, \enquote{Effects of
  self-and cross-phase modulation on photon purity for four-wave-mixing photon
  pair sources,} {\protect\JournalTitle{Physical Review A}} \textbf{92}, 053849
  (2015).

\bibitem{helt2015spontaneous}
L.~Helt, M.~Steel, and J.~Sipe, \enquote{Spontaneous parametric downconversion
  in waveguides: what's loss got to do with it?} {\protect\JournalTitle{New
  Journal of Physics}} \textbf{17}, 013055 (2015).

\bibitem{koefoed2019complete}
J.~G. Koefoed and K.~Rottwitt, \enquote{Complete evolution equation for the
  joint amplitude in photon-pair generation through spontaneous four-wave
  mixing,} {\protect\JournalTitle{Physical Review A}} \textbf{100}, 063813
  (2019).

\bibitem{koefoed2017spectrally}
J.~G. Koefoed, S.~M. Friis, J.~B. Christensen, and K.~Rottwitt,
  \enquote{Spectrally pure heralded single photons by spontaneous four-wave
  mixing in a fiber: reducing impact of dispersion fluctuations,}
  {\protect\JournalTitle{Optics express}} \textbf{25}, 20835--20849 (2017).

\bibitem{selvaraja2009subnanometer}
S.~K. Selvaraja, W.~Bogaerts, P.~Dumon, D.~Van~Thourhout, and R.~Baets,
  \enquote{Subnanometer linewidth uniformity in silicon nanophotonic waveguide
  devices using cmos fabrication technology,} {\protect\JournalTitle{IEEE
  Journal of Selected Topics in Quantum Electronics}} \textbf{16}, 316--324
  (2009).

\bibitem{waqas2018cascaded}
A.~Waqas, D.~Melati, and A.~Melloni, \enquote{Cascaded mach--zehnder
  architectures for photonic integrated delay lines,}
  {\protect\JournalTitle{IEEE Photonics Technology Letters}} \textbf{30},
  1830--1833 (2018).

\bibitem{vernon2015strongly}
Z.~Vernon and J.~Sipe, \enquote{Strongly driven nonlinear quantum optics in
  microring resonators,} {\protect\JournalTitle{Physical Review A}}
  \textbf{92}, 033840 (2015).

\bibitem{quesada2021beyond}
N.~Quesada, L.~Helt, M.~Menotti, M.~Liscidini, and J.~Sipe, \enquote{Beyond
  photon pairs: Nonlinear quantum photonics in the high-gain regime,}
  {\protect\JournalTitle{arXiv preprint arXiv:2110.04340}}  (2021).

\bibitem{Lumerical}
\url{www.ansys.com}.

\end{thebibliography}

\end{document}